\documentclass{aa}
\usepackage{graphicx}
\usepackage{txfonts}
\usepackage[utf8]{inputenc}
\usepackage[T1]{fontenc}
\usepackage{microtype}
\usepackage{xcolor}
\usepackage{multirow}
\usepackage{array}
\usepackage{amsmath}
\usepackage{amssymb}
\usepackage{newunicodechar}
\usepackage{ifthen}
\usepackage{placeins}
\usepackage{dblfloatfix}
\usepackage{capt-of}
\usepackage[
    breaklinks=true,
    hidelinks
]{hyperref}

\hypersetup{
    pdftitle={PhySR: Physics-Informed Neural Network for Super-Resolution Reconstruction in Radio Synthesis Imaging},
    pdfauthor={H. K. Yang and L. Zhang}
}
\newunicodechar{×}{\ensuremath{\times}}
\newunicodechar{↑}{\ensuremath{\uparrow}}
\newunicodechar{√}{\ensuremath{\checkmark}}
\newunicodechar{⋅}{\ensuremath{\cdot}}

\newcommand{\insertwidefigure}[5]{%
  \begin{figure*}[!tb]
  \centering
  \IfFileExists{#1}{%
    \includegraphics[width=#2,height=0.80\textheight,keepaspectratio]{#1}%
  }{%
    \parbox[c][#3][c]{0.92\textwidth}{\centering\mbox{}}%
  }
  \caption{#4}
  \label{#5}
  \end{figure*}
}

\newif\ifdraftreferences
\draftreferencesfalse

\begin{document}

\title{PhySR: Physics-Informed Neural Network for Super-Resolution Reconstruction in Radio Synthesis Imaging}
\titlerunning{PhySR for Radio Synthesis Imaging}
\authorrunning{Yang et al.}


\author{
H. K. Yang\inst{1}\email{gs.hkyang24@gzu.edu.cn}
\and
L. Zhang\inst{1}\corrauth{lizhang.science@gmail.com}
\and
M. Zhang\inst{2,3}\email{zhangm@xao.ac.cn}
\and
Y. F. Huang\inst{4,5}\email{hyf@nju.edu.cn}
}

\institute{
Guizhou Provincial Laboratory of Big Data,
College of Big Data and Information Engineering,
Guizhou University,
Guiyang 550025, China
\and
Xinjiang Astronomical Observatory,
Chinese Academy of Sciences,
Urumqi 830011, PR China
\and
Key Laboratory of Radio Astronomy,
Chinese Academy of Sciences,
Urumqi 830011, PR China
\and
School of Astronomy and Space Science,
Nanjing University,
Nanjing 210023, China
\and
Key Laboratory of Modern Astronomy and Astrophysics
(Nanjing University), Ministry of Education,
Nanjing 210023, China
}

\date{}
\abstract
{Radio telescope arrays are constrained by physical conditions such as the
number of antennas and the baseline distribution, and their spatial-frequency
sampling is incomplete, resulting in limited resolution of the observed images
and in blurring, distortion, and the loss of small-scale structures caused by
coupling between the primary and synthesized beams, thereby posing challenges
to the precise recovery of astrophysical information. Existing general-purpose
model-driven methods remove coupling effects sequentially, causing error
accumulation, but cannot address limited imaging resolution. Data-driven methods
lack explicit physical constraints, compromising conformity with radio
interferometric imaging physics.}
{We therefore propose PhySR, an end-to-end physics-informed neural network
(PINN), to improve reconstruction quality and physical consistency.}
{The model uses a U-Net as its backbone and deeply integrates data-driven
learning with physical constraints, enabling a direct mapping from
low-resolution dirty images to high-fidelity reconstructed images. It
incorporates physical priors from radio interferometric imaging, progressively
increases feature resolution through a dynamic cascaded upsampling module, and
combines a multiscale feature residual module to compensate for high-frequency
textures and local details, thereby achieving the precise recovery of
small-scale and multiscale structures. As the first in a series on PINN-based
astronomical super-resolution under observational effects, this study evaluates
PhySR on simulated SKA-Mid data.}
{At $4\times$, PhySR achieved 44.65~dB PSNR, 0.9940 SSIM, and 0.0065 RMSE.
Compared with existing methods, PSNR and SSIM improved by approximately
13.23~dB and 0.3760; compared with mainstream deep learning models, they
improved by 6.50~dB and 0.0682, while RMSE decreased by 0.0069. PhySR remained
stable at $2\times$ and $8\times$, with low observation-domain consistency
errors. Observation-domain degradation preserved consistency with original
observations, confirming advantages in coupling-effect removal, small-scale
structure recovery, and physical consistency.}
{Combining physical constraints and data-driven learning, PhySR achieves
high-fidelity reconstruction without high-resolution labels, providing an
efficient solution for SKA and next-generation radio telescope imaging.}

\keywords{
radio continuum: general --
methods: numerical --
techniques: image processing --
techniques: interferometric --
techniques: high angular resolution
}

\maketitle
\nolinenumbers
\raggedbottom
\section{Introduction}\label{sec:introduction}

The Square Kilometre Array (SKA), the world's largest and most sensitive major scientific facility currently under construction, will usher cosmology and radio astronomy into a new era characterized by both precision observations and discovery-driven research \citep{article1}. The SKA consists of two components: the low-frequency array (SKA-Low) and the mid-frequency array (SKA-Mid). SKA-Low covers the frequency range from 50 MHz to 350 MHz and primarily focuses on the early evolution of the Universe and the detection of low-frequency radio sources, whereas SKA-Mid, with its broad frequency coverage from 350 MHz to 15.4 GHz, serves as a key platform for probing the distribution of neutral hydrogen (H I), searching for pulsars, detecting fast radio bursts, and measuring baryon acoustic oscillations (BAO). With the completion and operation of the SKA, major breakthroughs are expected in research on a range of fundamental astrophysical questions, including primordial fluctuations in the Universe, tests of theories of gravity, and the nature of dark matter and dark energy, thereby advancing our understanding of the Universe to a new level \citep{article2}.

However, as the world's largest aperture-synthesis radio telescope system, the SKA faces a series of challenges in the imaging process. The limited number of antennas gives rise to the synthesized-beam effect, whose non-negligible sidelobes and broad main lobe introduce artifacts \citep{article3}. The telescope gain varies with direction, and the primary-beam effect is characterized by a gradual attenuation in gain from the phase center toward the edge \citep{article4}. In addition, limitations in the baseline lengths of radio interferometers restrict the angular resolution of the observed images, leaving image details poorly resolved. The combined effects of these three factors severely degrade image quality \citep{article5}. To address the image artifacts and blurring caused by the coupling between the primary beam and the synthesized beam, image reconstruction methods are required to reconstruct the true sky brightness distribution from the raw data acquired by the telescope \citep{article6}. To address the limited image resolution and loss of small-scale structures caused by incomplete spatial-frequency sampling, super-resolution reconstruction methods are commonly employed to improve image clarity \citep{article7}.

The most widely used methods for eliminating the synthesized-beam effect are CLEAN-type algorithms. The classical CLEAN deconvolution algorithm  \citep{article8} progressively removes the effects of the point spread function (PSF) from the observed data through iterative optimization, but its performance in processing extended sources is not satisfactory. Cornwell's Multiscale CLEAN (MS-CLEAN) was the first to model the sky brightness as a superposition of multiscale emission components and substantially improved the signal recovery performance for extended sources by simultaneously searching over multiple predefined scales rather than processing them sequentially \citep{article9}; however, the algorithm still relies on empirically predefined basis-function shapes and scale ranges, making it difficult to accommodate the scale uncertainty of complex astronomical sources. To address this limitation, Bhatnagar et al. proposed the ASP-CLEAN method, which dynamically determines the optimal scale through iterative fitting using an adaptive-scale pixel decomposition strategy, without requiring manual presetting, thereby further improving the reconstruction accuracy of multiscale structures \citep{article10}. Building on this work, Zhang et al. approximated the point spread function (PSF) with a single Gaussian function and analytically computed the model components, improving the computational efficiency of the adaptive-scale framework by more than a factor of 20 while maintaining the imaging performance \citep{article11}. In addition, the team tightly integrated the adaptive-scale sky model with wide-field imaging techniques, thereby achieving, for the first time, adaptive-scale reconstruction for wide-field imaging that accounts for the synthesized-beam effect and effectively eliminating this effect; this was the first validation of an adaptive-scale method using real observational data \citep{article12}. The primary-beam effect is also an important factor affecting image reconstruction accuracy, and traditional correction methods mainly include image-domain primary beam correction (PBC) \citep{article13} and the visibility-domain A-projection technique \citep{article14}, both of which require the establishment of a physical model of the primary beam before model-based bias correction is applied to the observational data. However, the actual primary beam exhibits azimuthal asymmetry, frequency-dependent sidelobe structures, and ripple effects, while traditional primary beam models, such as Gaussian and Jinc functions, can fit only the basic profile of the main lobe and cannot accurately characterize its azimuthal deviations, far-sidelobe distribution, or frequency-dependent variations \citep{article15}, thereby limiting the accuracy of such correction methods and rendering them unable to meet current requirements for high-dynamic-range and high-precision imaging \citep{article16}. Various super-resolution reconstruction methods are currently available for improving the resolution of astronomical images \citep{article17}. Among these, image-domain interpolation methods constitute the most direct means of improving resolution, with common approaches including nearest-neighbor interpolation and bilinear interpolation; these methods enlarge the image by estimating pixel intensities from neighboring pixels, but essentially perform only numerical interpolation between existing pixels \citep{article18}. In addition to interpolation methods, traditional super-resolution reconstruction methods based on sparse modeling have also been applied to radio synthesis imaging; these methods typically reconstruct a higher-resolution sky brightness distribution under incomplete spatial-frequency sampling by introducing prior constraints such as sparsity and regularization. However, such methods rely on prior assumptions and parameter settings and are prone to detail loss, over-smoothing, and the spurious recovery of high-frequency details under insufficient sampling or noise contamination \citep{article7}. Traditional super-resolution reconstruction methods generally face the challenges of high-frequency information loss and the spurious recovery of high-frequency details and therefore cannot meet the imaging requirements for the high-fidelity and high-precision reconstruction of fine structures at small scales.

In recent years, deep learning methods have been widely applied to the super-resolution reconstruction of astronomical images, leading to significant advances \citep{article19,article20,article21}. Among these, the convolutional neural network (CNN) is the most commonly used deep learning framework. The DBSRCNN network proposed by Albluwi et al. \citep{article22} integrated feature-extraction, feature-enhancement, and feature-concatenation layers into a CNN, thereby significantly improving the recovery of high-frequency image details, and is optimized for scenarios with both known and unknown blur levels; however, the model relies heavily on labeled high-resolution image data, which limits its generalization capability. Owing to their local feature-extraction mechanism, CNN-based deep learning methods require high-resolution images to provide sufficient detail and structural information. Therefore, these methods often rely heavily on high-resolution image data. However, high-resolution astronomical images are extremely difficult to obtain, resulting in scarce training data and further limiting the generalization capability of the models. Through its symmetric encoder--decoder structure and skip connections, U-Net \citep{article23} significantly improves the restoration quality of low-resolution images and can be trained with only a small amount of labeled data. Although U-Net can reduce the dependence on high-resolution data, it still relies on a purely data-driven learning approach and lacks physical consistency, which may cause the model's restoration results not to conform to established physical laws. The emergence of the physics-informed neural network (PINN) \citep{article24} addresses this limitation by encoding physical laws, represented by partial differential equations (PDEs), as physical residual terms in the loss function and combining them with a data-fitting loss in a dual-objective optimization mechanism, thereby ensuring the model's generalization capability with only a small amount of labeled data \citep{article24}. By combining data-driven learning with constraints imposed by physical laws, the PINN not only ensures that the restoration results conform to physical laws but also further reduces the model's dependence on high-resolution data.

However, no study in radio astronomy has yet investigated super-resolution reconstruction in the presence of observational effects. The aim of this work is to achieve super-resolution reconstruction without high-resolution labels while ensuring that the reconstructed results conform to the physical principles of radio interferometry. Therefore, using U-Net as the backbone, we propose a novel super-resolution image reconstruction method, PhySR (Physics-Informed Super-Resolution Network), which tightly integrates PINN-based physical constraints with the strengths of data-driven learning. While fully leveraging U-Net's powerful feature-extraction and detail-recovery capabilities for low-resolution images, PhySR incorporates the physical principles of radio interferometry into the loss constraints; it can simultaneously perform image restoration and super-resolution reconstruction, producing super-resolved images that conform to physical laws.

\section{Radio Synthesis Imaging}\label{sec:imaging}

Radio synthesis imaging is a core technique in radio astronomy for overcoming the angular-resolution limitations of a single antenna and enabling high-angular-resolution observations \citep{article5}. According to the Van Cittert-Zernike theorem \citep{article25}, under ideal continuous-sampling conditions, the visibility data in the spatial-frequency domain and the sky brightness distribution in the image domain constitute a two-dimensional Fourier-transform pair, given by

\begin{equation}
V(u,v) = \iint_{}^{}I_{true}(l,m)e^{- 2\pi i(ul + vm)}dldm
\label{eq:01}
\end{equation}

Where \(I_{true}(l,m)\) denotes the true sky brightness distribution, and \((l,m)\) denote the sky coordinates on the tangent plane, measured in radians; \(\text{V}\text{(}\text{u}\text{,}\text{v}\text{)}\) denotes the visibility function in the spatial-frequency domain, and \(\text{(}\text{u}\text{,}\text{v}\text{)}\) denote the spatial-frequency coordinates, which are determined by the projection of the antenna baseline onto the plane perpendicular to the source direction and by the observing wavelength.

In practical radio interferometric observations, the true sky brightness is not measured directly under ideal conditions; instead, it is first modulated by the antenna's direction-dependent gain pattern, that is, by the primary-beam effect. Let \(\text{B}\text{(}\text{l}\text{,}\text{m}\text{)}\) denote the primary beam response function; the sky brightness distribution after primary-beam modulation is then given by

\begin{equation}
I_{beam}(l,m) = I_{true}(l,m) \odot B(l,m)
\label{eq:02}
\end{equation}

where $\odot$ denotes element-wise multiplication. Upon substituting the above expression into Eq.~(\ref{eq:01}), the corresponding ideal observed visibility is given by

\begin{equation}
V_{ideal}(u,v) = \iint_{}^{}\left\lbrack I_{true}(l,m) \odot B(l,m) \right\rbrack e^{- 2\pi i(ul + vm)}dldm
\label{eq:03}
\end{equation}

Because the number of antennas in a practical interferometric array is finite and the baseline lengths and their distribution are constrained by physical conditions, visibility measurements can be obtained only at discrete sampling points in the (uv) plane, rather than continuously across the entire spatial-frequency domain. Let (S(u,v)) denote the sampling function in the (uv) plane; the observed visibility is then given by

\begin{equation}
V_{obs}(u,v) = S(u,v)V_{ideal}(u,v)
\label{eq:04}
\end{equation}

Substitution of Eq.~(\ref{eq:02}) into the preceding equation, followed by a two-dimensional inverse Fourier transform, yields the image-domain observation given by

\begin{equation}
I_{dirty}(l,m) = \mathcal{F}^{- 1}\{ V_{obs}(u,v)\}
\label{eq:05}
\end{equation}

According to the convolution theorem, the inverse Fourier transform of the sampling function \(\text{S}\text{(}\text{u}\text{,}\text{v}\text{)}\) corresponds to the synthesized beam in the image domain, denoted by \(P(l,m) = \mathcal{F}^{- 1}\{ S(u,v)\}\) Therefore, the dirty image obtained from actual observations is given by

\begin{equation}
I_{dirty}(l,m) = \left\lbrack I_{true}(l,m) \odot B(l,m) \right\rbrack \ast P(l,m)
\label{eq:06}
\end{equation}

As can be seen from Eq.~(\ref{eq:06}), the dirty image obtained from actual observations is jointly affected by the primary-beam and synthesized-beam effects. In the image domain, the primary-beam effect mainly manifests as a spatially nonuniform gain distribution, with the gain gradually decreasing as the distance from the phase center increases. The synthesized-beam effect mainly manifests as blurring and artifacts induced by the main-lobe and sidelobe responses. In actual observations, these two effects are coupled, further exacerbating image blurring. It should also be noted that the incomplete sampling of the continuous \(\text{uv}\) plane by the sampling function \(\text{S}\text{(}\text{u}\text{,}\text{v}\text{)}\) is, in essence, a form of undersampling, thereby affecting the resolution of the observed image.

Existing general-purpose methods can eliminate most of the primary-beam and synthesized-beam effects in observed images, but residual effects remain in the restoration results. Meanwhile, the sequential processing adopted by these methods causes errors to accumulate and amplify, ultimately degrading the quality of the restored images. It should be noted that existing general-purpose methods cannot directly reconstruct super-resolved images from dirty images.

\section{Motivation}\label{sec:motivation}

The super-resolution reconstruction of astronomical images is not simply a matter of image enlargement; rather, under the constraints imposed by the observing instrument, it is a complex inverse problem arising jointly from the coupling effects between the primary beam and the synthesized beam and from limited imaging resolution. Factors such as the direction-dependent antenna response and the baseline distribution lead to structural blurring, complex artifacts, and the loss of small-scale details in low-resolution dirty images. Therefore, a reconstruction model must not only increase the spatial resolution of the image but also eliminate the coupling effects introduced during the imaging process and ensure that the reconstructed results are physically consistent with the radio synthesis imaging process.

However, existing general-purpose methods have difficulty addressing the issues described above. Although purely data-driven deep learning methods for super-resolution have strong nonlinear mapping and image restoration capabilities, they still exhibit clear limitations in astronomical image reconstruction tasks (Fig.~\ref{fig:principle}). Most supervised deep learning methods require paired low-resolution images and high-resolution labels for training, whereas ground-truth high-resolution images are generally difficult to obtain directly from real radio observations, thereby constraining model training and generalization. Moreover, conventional deep learning methods rely primarily on learning the data distribution and lack explicit constraints on imaging physics, including the primary beam response, PSF convolution, and observation-domain consistency, and may therefore generate visually sharp but spurious details that are inconsistent with the actual observational process, thereby reducing the physical reliability of the reconstruction results \citep{article26}.

To address the limitations of existing general-purpose methods and data-driven deep learning methods for super-resolution, this study introduces physics-informed constraints and incorporates the primary beam response, PSF convolution, and observation-domain consistency into the super-resolution reconstruction process to develop a physics-informed super-resolution reconstruction model. While requiring no high-resolution labels, the model constrains the reconstruction result so that, after being degraded back to the observation domain, it can adequately account for the original low-resolution dirty image. Moreover, this study is the first to integrate the removal of these effects with super-resolution reconstruction for astronomical image processing.

\insertwidefigure{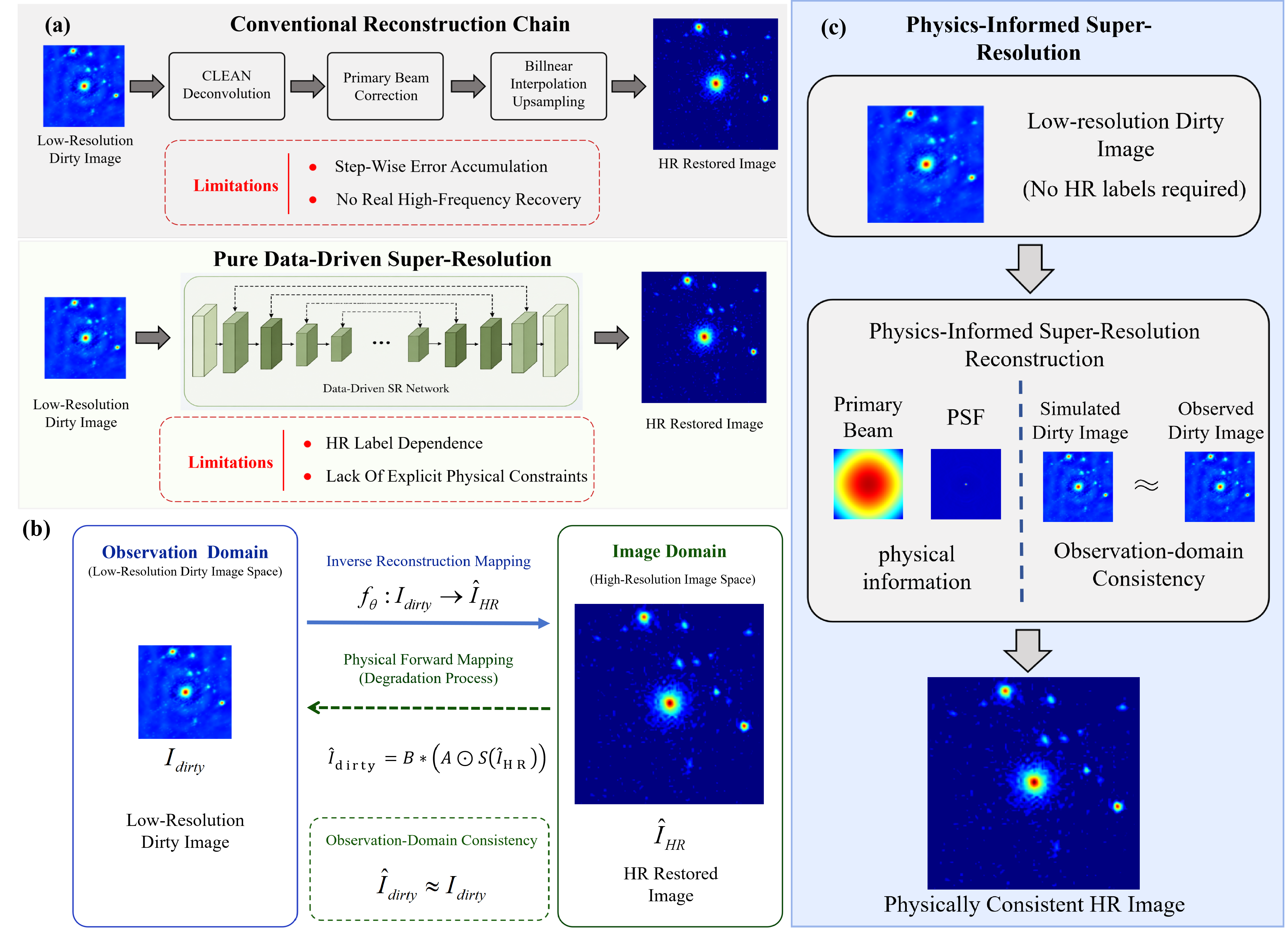}{0.94\textwidth}{7.5cm}{Principle of the PINN-based super-resolution reconstruction method. (a) Limitations of existing general-purpose methods and purely data-driven super-resolution methods. Existing general-purpose methods generally employ a sequential pipeline comprising CLEAN deconvolution, primary beam correction, and interpolation-based upsampling, which is prone to error accumulation and has difficulty recovering missing high-frequency information, whereas purely data-driven methods can directly learn the mapping from low-resolution dirty images to high-resolution images but generally require high-resolution labels and lack explicit physical constraints. (b) Physical mapping between the observation and image domains. The model learns an inverse reconstruction mapping to recover a high-resolution image from a low-resolution dirty image; meanwhile, a physical forward mapping is constructed through primary-beam modulation, PSF convolution, and scale mapping, so that the reconstructed image remains consistent with the original dirty image after being degraded back to the observation domain. (c) Physics-informed super-resolution reconstruction framework proposed in this work. Without relying on high-resolution labels, the method incorporates primary-beam, PSF, and observation-domain consistency constraints, ultimately yielding a high-resolution reconstructed image that conforms to the physical process of radio synthesis imaging.}{fig:principle}

\section{Methods}\label{sec:methods}
Existing image reconstruction and super-resolution methods cannot completely eliminate the coupling effects between the primary beam and the synthesized beam, while imaging resolution remains limited by the telescope baseline length; moreover, existing deep learning models lack explicit physical constraints on PSF convolution and the primary beam response, resulting in limited generalization capability when high-resolution data are scarce. To overcome these limitations, we propose PhySR (Physics-Informed Super-Resolution Network), a physics-constrained super-resolution reconstruction network with dynamic upsampling. This section provides an overview of the overall PhySR framework and then describes in detail its key components, including the dynamic cascaded upsampling module, the multiscale feature residual module, and the loss function.

\subsection{PhySR Network Model}\label{sec:network}

PhySR is a physics-informed super-resolution network based on a U-Net encoder--decoder architecture that incorporates physical priors from radio synthesis imaging, including PSF convolution and the primary beam response, to enhance the recovery of small-scale structures, faint-source signals, and edge contours and to eliminate artifacts caused by coupling effects. Through progressive resolution enhancement and cross-scale feature fusion, PhySR achieves the accurate recovery of small-scale structures and multiscale morphological features, ultimately establishing an end-to-end multilevel feature reconstruction mechanism, as shown in Fig.~\ref{fig:architecture}.

\insertwidefigure{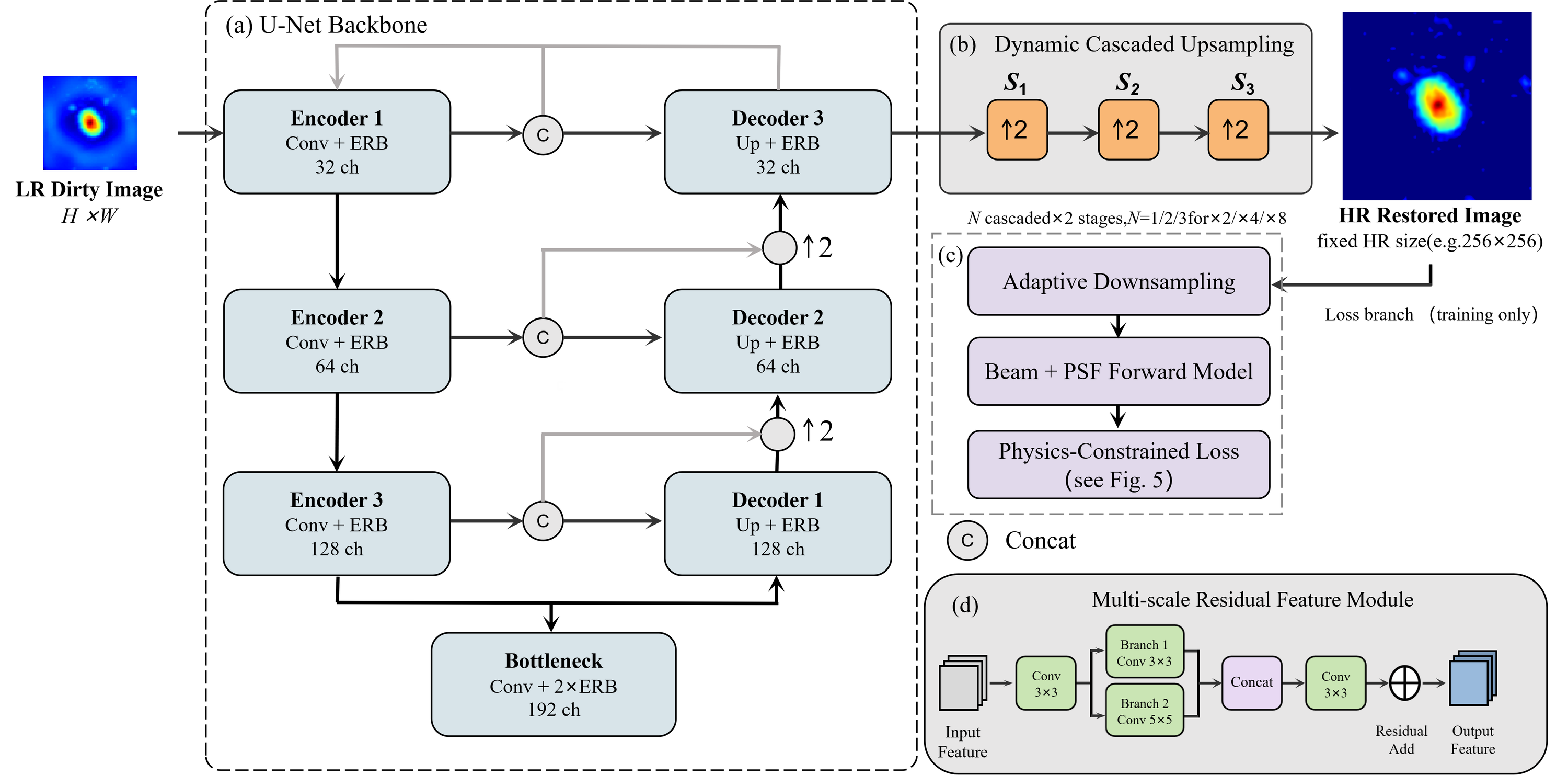}{0.94\textwidth}{6.0cm}{Architecture of the PhySR network. The leftmost and rightmost images represent the input low-resolution dirty image and the output high-resolution reconstructed image, respectively. (a) Three-level U-Net encoder--decoder backbone, comprising three encoder blocks, one bottleneck block, and three decoder blocks. Each encoder block consists of a convolutional layer and an enhanced residual block (ERB), whereas each decoder block consists of an upsampling layer and an ERB; the numbers below the blocks indicate the numbers of feature channels, the symbol C denotes feature concatenation through skip connections, and ↑2 denotes 2× upsampling between adjacent decoder stages. (b) Dynamic cascaded upsampling module, comprising three 2× upsampling stages, S1, S2, and S3; the number of cascaded stages, N, is selected according to the target super-resolution factor, with N = 1, 2, and 3 corresponding to 2×, 4×, and 8× super-resolution reconstruction, respectively, ultimately generating a high-resolution reconstructed image at a fixed resolution. (c) Physics-constraint branch used only during training, comprising adaptive downsampling, the Beam + PSF forward model, and the physics-constrained loss illustrated in Fig.~\ref{fig:forward}. (d) Multiscale feature residual module, in which the input feature first passes through a 3 × 3 convolution, after which parallel 3 × 3 and 5 × 5 convolutional branches extract multiscale features; the branch features are then concatenated, fused by a 3 × 3 convolution, and added to the input feature through a residual connection to produce the output feature.}{fig:architecture}

The PhySR architecture consists primarily of three components: a feature encoding--decoding backbone, a dynamic cascaded upsampling module, and a multiscale feature residual module. After shallow convolutional mapping, the input low-resolution dirty image enters the encoder, where features are extracted sequentially through multiple convolutional layers and enhanced residual blocks; over three downsampling stages, multiscale representations are progressively learned, ranging from local textures and edge information to high-level structural semantics, while the features from each encoder stage are transferred to the corresponding decoder stage through skip connections to preserve low-level detail priors. At the bottleneck, enhanced residual blocks further aggregate deep features to strengthen the global modeling of the main structures, faint-source responses, and background distribution. The decoder progressively restores the spatial resolution through successive upsampling and fuses the resulting features with those from the corresponding encoder stages to supplement detailed information and reduce the blurring and artifacts caused by coupling effects. Once the decoder features at the input resolution have been obtained, they are further processed by the dynamic cascaded upsampling module and the multiscale feature residual module, with the former progressively increasing the feature resolution in steps of 2× and the latter compensating for high-frequency textures and small-scale structural details through residual branches; the target high-resolution reconstructed image is ultimately generated by the output convolutional layer, thereby achieving end-to-end physics-informed super-resolution reconstruction.

\subsubsection{Dynamic Cascaded Upsampling Module}\label{sec:cascaded}

The dynamic cascaded upsampling module progressively enlarges the low-resolution features produced by the decoder to the target high resolution. At each upsampling stage, it maintains the continuity of information across scales, allowing local textures and structural details in low-level features to be progressively propagated into the high-resolution feature space. As shown in Fig.~\ref{fig:cascaded}, each upsampling stage applies a 2× scale increase to ensure the progressive expansion of the feature space while fully preserving information from the encoder.

\insertwidefigure{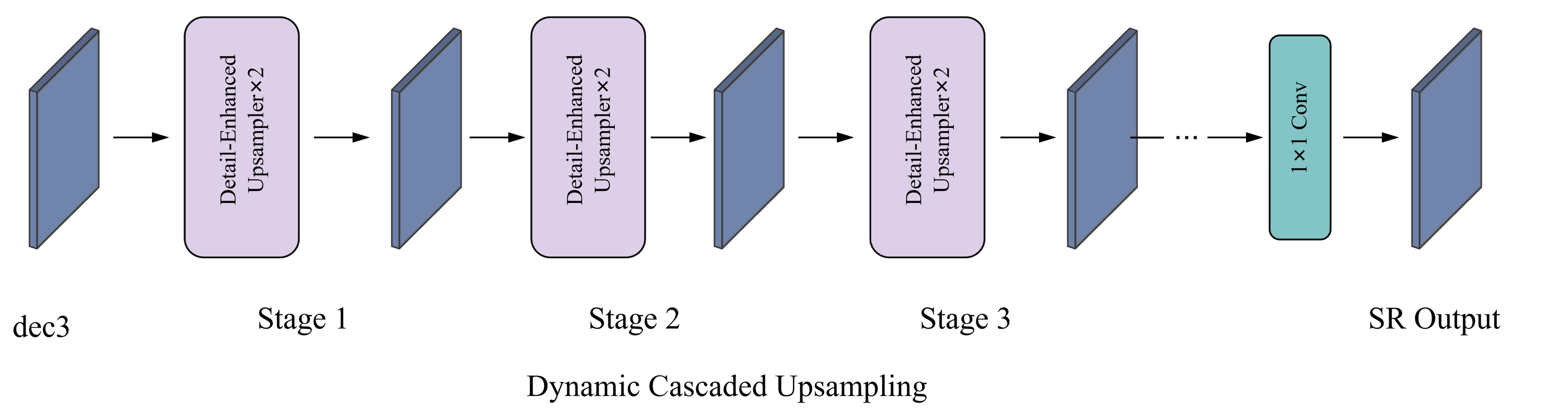}{0.94\textwidth}{4.0cm}{Dynamic cascaded upsampling module.}{fig:cascaded}

Let the target super-resolution factor be \(s\), and let the number of cascaded stages be \(L = \log_{2}s\); the upsampled feature at level \(l\) is then given by

\begin{equation}
F_{up}^{(l)} = \mathcal{U}_{2}(F_{up}^{(l - 1)}),\quad l = 1,2,\ldots,L
\label{eq:07}
\end{equation}

where \(\mathcal{U}_{2}( \cdot )\) denotes a \(\text{2}\text{×}\) upsampling operator, typically implemented using bilinear interpolation or convolution-based upsampling. Through progressive upsampling, the feature space is expanded from the original decoder resolution \(\text{H}\text{×}\text{W}\) to the target high resolution \(\text{sH}\text{×}\text{sW}\).

After each upsampling stage, the module also preserves the continuity and stability of information across scales, allowing details from low-level features to be progressively integrated into higher-resolution representations. After \(L\) upsampling stages, the target-scale feature \(\text{F}_{\text{up}}^{\text{(}\text{L}\text{)}}\), which serves as the high-resolution input to the subsequent multiscale feature residual module, is given by

\begin{equation}
F_{msr\_ in} = F_{up}^{(L)}
\label{eq:08}
\end{equation}

where \(F_{msr\_ in}\) denotes the input to the multiscale feature residual module and is further processed to compensate for high-frequency details and enhance local textures.

\subsubsection{Multiscale Feature Residual Module}\label{sec:msr}

The multiscale feature residual module compensates for high-frequency details and enhances local textures in the high-resolution features produced by the dynamic cascaded upsampling module. It extracts local information with different receptive fields through parallel convolutional branches and fuses multiscale information through feature concatenation and residual connections, thereby improving the reconstruction accuracy of small-scale structures and edge textures. The module is illustrated in Fig.~\ref{fig:msr}.

\insertwidefigure{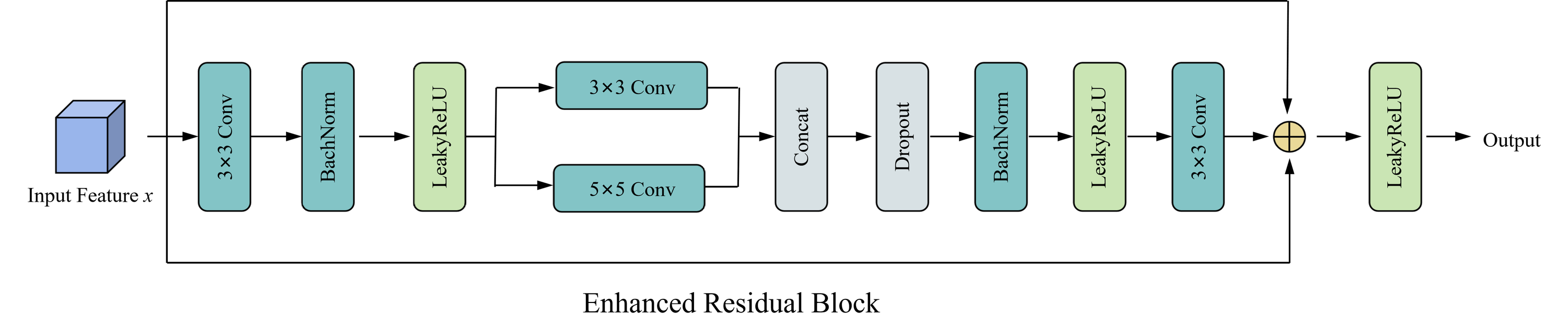}{0.94\textwidth}{4.0cm}{Multiscale feature residual module.}{fig:msr}

Let \(F_{msr\_ in}\) denote the output feature of the dynamic cascaded upsampling module; the residual features computed by the two convolutional branches of the multiscale feature residual module are then given by

\begin{equation}
R_{3 \times 3} = \mathcal{R}_{3 \times 3}(F_{msr\_ in}),\quad R_{5 \times 5} = \mathcal{R}_{5 \times 5}(F_{msr\_ in})
\label{eq:09}
\end{equation}

where \(R_{3 \times 3}( \cdot )\) and \(R_{5 \times 5}( \cdot )\) denote the 3×3 and 5×5 convolutional branches, respectively; the output feature of the module obtained through residual fusion is given by

\begin{equation}
F_{msr\_ out} = F_{msr\_ in} + R_{3 \times 3} + R_{5 \times 5}
\label{eq:10}
\end{equation}

Finally, a convolutional mapping is applied to the enhanced feature \(F_{msr\_ out}\) to produce the high-resolution reconstructed image, given by

\begin{equation}
I_{HR} = {Conv}_{out}(F_{MSR\_ out})
\label{eq:11}
\end{equation}

\subsection{Loss Function}\label{sec:loss}

The central aim of the loss-function design is to incorporate the physical principles of radio synthesis imaging and combine data fitting with regularization to achieve high-fidelity radio imaging without high-resolution labels. By encoding the physical mechanisms as differentiable loss terms, this design constrains the model output to maintain physical consistency while balancing detail recovery against noise and artifact suppression, thereby providing a physics-driven optimization strategy for addressing the ill-posedness of the inverse problem in radio synthesis imaging. The total loss function is defined as

\begin{equation}
\mathcal{L}_{total} = \omega_{phy} \cdot \mathcal{L}_{phy} + \omega_{ssim} \cdot \mathcal{L}_{ssim} + \omega_{mse} \cdot \mathcal{L}_{mse} + \omega_{reg} \cdot \left( \mathcal{L}_{tv} + \mathcal{L}_{artifact} \right)
\label{eq:12}
\end{equation}

where \(\omega_{\text{phy}}\), \(\omega_{\text{ssim}}\), \(\omega_{\text{mse}}\), and \(\omega_{\text{reg}}\) are the weighting coefficients for the physical loss, structural similarity loss, mean squared error loss, and regularization loss, respectively; \(\mathcal{L}_{total}\) denotes the total loss function; \(\mathcal{L}_{phy}\) denotes the core physical-constraint loss incorporating the primary-beam and synthesized-beam forward model; \(\mathcal{L}_{ssim}\) and \(\mathcal{L}_{mse}\) constrain the reconstruction result in terms of structural consistency and pixel intensity, respectively; \(\mathcal{L}_{tv}\) suppresses noise amplification while preserving background smoothness; and \(\mathcal{L}_{artifact}\) suppresses structured artifacts introduced by the sidelobes.

\insertwidefigure{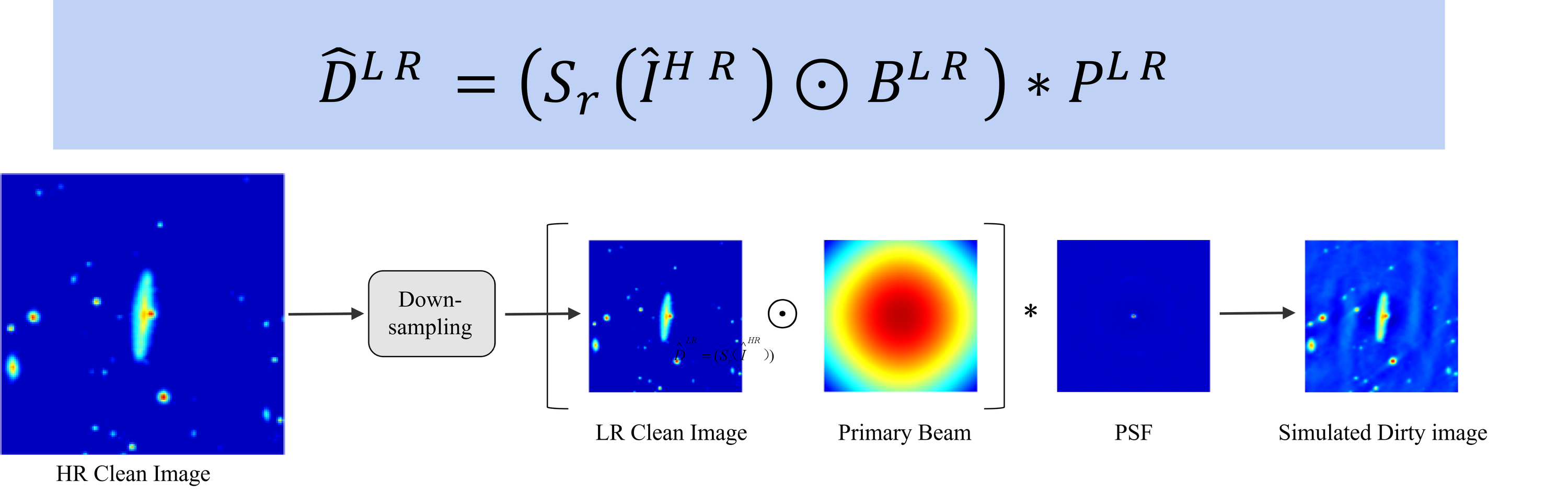}{0.94\textwidth}{4.5cm}{Schematic of the physical forward model for radio synthesis imaging. The high-resolution reconstructed image is first downsampled to the observation scale, then weighted by the primary beam response and convolved with the PSF to generate a simulated dirty image.}{fig:forward}

\subsubsection{Physical Loss}\label{sec:physical_loss}

To achieve physically consistent super-resolution reconstruction without high-resolution labels, this work explicitly incorporates the degradation mechanism of radio synthesis imaging into the training process, constructs a forward model comprising a primary-beam weighting operator, a synthesized-beam convolution operator, and a scale-mapping operator (downsampling), and uses this model together with an observation-domain consistency constraint to define the physical loss, as shown in Fig.~\ref{fig:forward}.

During the reconstruction process, let the input be the observed low-resolution dirty image \(\text{D}^{\text{LR}}\), and let the network output be the high-resolution reconstructed image \({\widehat{\text{I}}}^{\text{HR}}\). Because direct supervision from the high-resolution ground truth \(\text{I}^{\text{HR}}\) is unavailable during training, this work maps \({\widehat{\text{I}}}^{\text{HR}}\) back to the original observation scale and constrains the resulting simulated dirty image to remain consistent with the observed low-resolution dirty image, thereby imposing an effective physical constraint on \({\widehat{\text{I}}}^{\text{HR}}\).

First, the high-resolution reconstructed image \({\widehat{\text{I}}}^{\text{HR}}\) is downsampled to the observation resolution using the scale-adaptive downsampling operator \(S_{r}( \cdot )\) corresponding to the super-resolution factor \(r\), such that \({\widehat{\text{I}}}^{\text{LR}}\text{=}\text{S}_{\text{r}}\text{(}{\widehat{\text{I}}}^{\text{HR}}\text{)}\), and is then aligned with the low-resolution clean reference image \(\text{I}^{\text{LR}}\) to constrain the morphology and intensity scale of the network output to remain consistent with those of the reference image at the observation scale, thereby avoiding global brightness shifts or structural distortions during super-resolution reconstruction. Meanwhile, to ensure that the reconstructed result conforms to the physical process of radio synthesis imaging, \({\widehat{\text{I}}}^{\text{HR}}\) is passed through the differentiable degradation model derived from Eq.~(\ref{eq:03}) to generate a simulated dirty image at the observation scale, given by

\begin{equation}
{\widehat{D}}^{LR} = \left( S_{r}({\widehat{I}}^{HR}) \odot B^{LR} \right) \ast P^{LR}
\label{eq:13}
\end{equation}

where \(\text{P}^{\text{LR}}\) denotes the synthesized beam (PSF), and \(\text{B}^{\text{LR}}\) denotes the primary beam response map. The discrepancy between \({\widehat{\text{D}}}^{\text{LR}}\) and the dirty image \(\text{D}^{\text{LR}}\) obtained from actual observations is then used as an observation-domain consistency constraint, thereby imposing a physical constraint that prevents the network from generating spurious high-frequency details and overfitting-induced textures that are inconsistent with the instrumental degradation process.

Accordingly, the physical loss is defined as

\begin{equation}
\mathcal{L}_{phy} = \|{\widehat{\mathbf{D}}}^{LR} - \mathbf{D}^{LR}\|_{1} + \alpha\|{\widehat{\mathbf{I}}}^{LR} - \mathbf{I}^{LR}\|_{1}
\label{eq:14}
\end{equation}

where \(\|{\widehat{\mathbf{D}}}^{LR} - \mathbf{D}^{LR}\|_{1}\) denotes the observation-domain physical-consistency term, \(\|{\widehat{\mathbf{I}}}^{LR} - \mathbf{I}^{LR}\|_{1}\) denotes the content-consistency term at the observation scale, and \(\alpha\) controls the relative weighting of the two terms in the joint optimization.

\subsubsection{Structural and Pixel-Level Losses and Regularization Constraints}\label{sec:other_losses}

Building on the physical constraint imposed on \({\widehat{\text{I}}}^{\text{HR}}\) by the physical loss \(\mathcal{L}_{phy}\) through observation-domain consistency, this work further introduces a structural similarity constraint, a pixel-domain fitting constraint, and regularization terms to formulate a multi-objective joint optimization that simultaneously preserves structural consistency, constrains pixel-intensity deviations, and suppresses noise and artifacts, thereby improving the stability and perceptual quality of super-resolution reconstruction in radio astronomy.

To impose stronger constraints on the morphology and local contrast of astronomical sources, we employ a structural similarity loss, given by

\begin{equation}
\mathcal{L}_{ssim} = 1 - SSIM({\widehat{I}}^{LR},I^{LR})
\label{eq:15}
\end{equation}

where \({\widehat{\text{I}}}^{\text{LR}}\text{=}\text{S}_{\text{r}}\text{(}{\widehat{\text{I}}}^{\text{HR}}\text{)}\), and the structural similarity loss characterizes similarity in terms of luminance, contrast, and structure \citep{article27}, helping to preserve point-source morphology, the contours of extended sources, and local structural consistency. Meanwhile, to provide a stable amplitude-fitting constraint and enforce consistency of the overall intensity scale, the mean squared error loss is introduced and given by

\begin{equation}
\mathcal{L}_{mse} = \left\| {\widehat{I}}^{LR} - I^{LR} \right\|_{2}^{2}
\label{eq:16}
\end{equation}

This term is more sensitive to global photometric deviations and complements \(\mathcal{L}_{ssim}\), thereby allowing the training process to balance structural fidelity and amplitude consistency.

Moreover, because super-resolution reconstruction can amplify high-frequency noise and cause local noise enhancement, the total variation regularization term is introduced to suppress abrupt local variations and preserve background smoothness and is defined as

\begin{equation}
\mathcal{L}_{tv} = \sum_{i,j}^{}\left( |{\widehat{I}}_{i + 1,j}^{HR} - {\widehat{I}}_{i,j}^{HR}| + |{\widehat{I}}_{i,j + 1}^{HR} - {\widehat{I}}_{i,j}^{HR}| \right)
\label{eq:17}
\end{equation}

To suppress structured artifacts caused by the sidelobes of the synthesized beam, we further impose a lightweight suppression constraint in the observation-residual domain. Let \({\widehat{D}}_{LR} = \mathcal{A}({\widehat{I}}^{HR})\) and \(R^{LR} = D^{LR} - {\widehat{D}}^{LR}\); the constraint is then given by

\begin{equation}
\mathcal{L}_{artifact} = \left\| \nabla_{x}R^{LR} \right\|_{1} + \left\| \nabla_{y}R^{LR} \right\|_{1}
\label{eq:18}
\end{equation}

Where \(\nabla_{x}\) and \(\nabla_{y}\) denote the horizontal and vertical gradient operators, respectively, used to suppress directional texture patterns and stripe-like structures in the residual, thereby reducing the sidelobe artifacts remaining in the reconstructed image.

\FloatBarrier

\section{Results}\label{sec:results}

This chapter systematically evaluates the reconstruction performance of PhySR on astronomical images. The experimental design covers three aspects: first, ablation experiments are conducted to assess the effects of the dynamic cascaded upsampling module and the multiscale feature residual module on the reconstruction results. Next, PhySR is compared quantitatively and qualitatively with existing general-purpose methods. Finally, a comprehensive comparison is conducted with mainstream deep learning methods for super-resolution. Notably, this work also introduces observation-domain consistency metrics, which quantify the reliability and consistency of the reconstruction results from the perspective of the physical observation process by degrading the high-resolution reconstructed images to the observation domain and comparing them with the original observed images, thereby validating the model's performance advantages and novelty across multiple dimensions.

\subsection{Dataset}\label{sec:dataset}

To evaluate PhySR's ability to eliminate the coupling effects between the primary and synthesized beams under multiple observing scenarios, as well as its super-resolution reconstruction performance without high-resolution labels, we selected 3000 real astronomical images from Galaxy Zoo as ideal sky models and simulated the telescope observing process using the interferometric imaging simulator OSKAR 2.7.6 \citep{article28} and the SKA mid-frequency array (ska1mid.tm) \citep{article29}, introducing the coupling effects between the primary and synthesized beams to generate dirty images with controllable degradation characteristics and ultimately constructing a dataset of 9000 clean--dirty image pairs at three resolutions---32×32, 64×64, and 128×128 pixels---with the key observing parameter settings listed in Table~\ref{tab:simulation_parameters}.

\begin{table}[!t]
\caption{Simulated observing parameters at different resolutions.}
\label{tab:simulation_parameters}
\centering
\small
\setlength{\tabcolsep}{3.5pt}
\renewcommand{\arraystretch}{1.12}

\begin{tabular}{lll}
\hline\hline
Resolution & Parameter & Value \\
\hline

\multirow{6}{*}{$32\times32$}
 & Cell size (arcsec) & 20.0 \\
 & Start frequency (Hz) & $5.0\times10^{9}$ \\
 & Mid-frequency array & \texttt{ska1mid.tm} \\
 & Observation length & 17:00:00.0 \\
 & Frequency increment (Hz) & $30\times10^{6}$ \\
 & Channel bandwidth (Hz) & $1\times10^{6}$ \\
\hline

\multirow{6}{*}{$64\times64$}
 & Cell size (arcsec) & 20.0 \\
 & Start frequency (Hz) & $3.0\times10^{9}$ \\
 & Mid-frequency array & \texttt{ska1mid.tm} \\
 & Observation length & 15:00:00.0 \\
 & Frequency increment (Hz) & $20\times10^{6}$ \\
 & Channel bandwidth (Hz) & $1\times10^{6}$ \\
\hline

\multirow{6}{*}{$128\times128$}
 & Cell size (arcsec) & 20.0 \\
 & Start frequency (Hz) & $9.5\times10^{8}$ \\
 & Mid-frequency array & \texttt{ska1mid.tm} \\
 & Observation length & 12:00:00.0 \\
 & Frequency increment (Hz) & $10\times10^{6}$ \\
 & Channel bandwidth (Hz) & $1\times10^{6}$ \\
\hline
\end{tabular}

\end{table}

For the 32×32 resolution, the observing parameters comprise a starting frequency of 5 GHz, an observing duration of 17 h, a frequency increment of 30 MHz, a cell size of 20 arcsec, and a channel bandwidth of 1 MHz, corresponding to the extreme-degradation setting for 8× super-resolution and used to evaluate the model's basic reconstruction capability and robustness under extreme degradation; for the 64×64 resolution, the observing parameters comprise a starting frequency of 3 GHz, an observing duration of 15 h, a frequency increment of 20 MHz, a cell size of 20 arcsec, and a channel bandwidth of 1 MHz, corresponding to the strong-degradation setting for 4× super-resolution and used to evaluate the model's ability to balance the recovery of large-scale structures and small-scale details; and for the 128×128 resolution, the observing parameters comprise a starting frequency of 950 MHz, an observing duration of 12 h, a frequency increment of 10 MHz, a cell size of 20 arcsec, and a channel bandwidth of 1 MHz, corresponding to the weak-degradation setting for 2× super-resolution and used to evaluate the model's ability to reconstruct high-frequency details and small-scale structures with high fidelity.

All datasets were independently divided into training, validation, and test sets at a ratio of 8:1:1; for each resolution, the training set contained 2400 image pairs, while the validation and test sets each contained 300 image pairs, and no supervision from high-resolution labels was introduced throughout training. These dataset splits were used to systematically evaluate the model's super-resolution reconstruction performance and generalization capability.

\subsection{Experimental Environment and Configuration}\label{sec:environment}

To ensure comparability among methods and the reproducibility of the experimental results, all experiments were conducted in the same software and hardware environment, and all methods included in the comparison were evaluated using exactly the same experimental settings. The hardware environment consisted of an Intel(R) Xeon(R) E5410 CPU operating at 2.33 GHz, 128 GB of RAM, and six NVIDIA GeForce RTX 2080 Ti GPUs, each equipped with 12 GB of GPU memory; the software environment consisted of the Ubuntu 18.04 operating system, with GPU-accelerated computation implemented using CUDA 11.4.
\subsection{Performance Metrics}\label{sec:metrics}

\subsubsection{Image-Domain Metrics}\label{sec:image_metrics}

To ensure the objectivity and rigor of the experimental results and comparability among algorithms, this work adopts three core quantitative evaluation metrics: peak signal-to-noise ratio (PSNR), structural similarity index (SSIM), and root mean square error (RMSE). These metrics provide unified and objective criteria for comparing algorithm performance from the three perspectives of numerical performance, visual perception, and reconstruction error.

\textbf{Peak signal-to-noise ratio (PSNR)}\citep{article30}\textbf{:} PSNR is an image-fidelity metric based on the pixel-wise mean squared error and is one of the most widely used fundamental
quantitative metrics in image quality assessment; it measures the overall
pixel-wise discrepancy between the predicted and ground-truth values. It is
given by

\begin{equation}
PSNR = 10 \cdot \log_{10}\left( \frac{MAX_{L}^{2}}{MSE} \right)
\label{eq:19}
\end{equation}

\textbf{Structural similarity index measure (SSIM)} \citep{article31}\textbf{:} SSIM comprehensively evaluates the structural consistency between two images in terms of luminance, contrast, and structure, thereby addressing the limitations of PSNR, which captures only the overall pixel-wise error and cannot characterize structural integrity or visual similarity, and serves as a core quantitative metric for assessing image structural fidelity and visual quality. It is given by

\begin{equation}
SSIM(x,y) = \frac{(2\mu_{x}\mu_{y} + C_{1})(2\sigma_{xy} + C_{2})}{(\mu_{x}^{2} + \mu_{y}^{2} + C_{1})(\sigma_{x}^{2} + \sigma_{y}^{2} + C_{2})}
\label{eq:20}
\end{equation}

\textbf{Root mean square error (RMSE)} \citep{article32}\textbf{:} RMSE is a core metric for measuring the overall level of pixel-wise error between a reconstructed image and the corresponding ground-truth image; it directly reflects the overall deviation of the reconstruction result from the reference and serves as a fundamental quantitative metric for evaluating an algorithm's error-suppression capability. It is given by

\begin{equation}
RMSE = \sqrt{\frac{1}{H \cdot W}\sum_{i=1}^{H}\sum_{j=1}^{W}\left(I_{true}(i,j)-I_{rec}(i,j)\right)^2}
\label{eq:21}
\end{equation}

\subsubsection{Observation-Domain Consistency Metrics}\label{sec:obs_metrics}

In addition to image-domain evaluation metrics such as PSNR, SSIM, and RMSE, this work further introduces observation-domain consistency metrics to assess the reliability of high-resolution reconstruction results from the perspective of the physical process of radio synthesis imaging. Image-domain metrics primarily measure the pixel-wise error and structural similarity between the reconstructed image and the reference image, but they cannot directly determine whether the high-resolution structures recovered by the model can be reasonably explained by the actual observational process. Therefore, this work degrades the high-resolution reconstructed image output by the model back to the original observation scale and compares it with the input dirty image to determine whether the reconstruction result satisfies the physical consistency jointly constrained by the primary beam response, PSF convolution, and observation-scale mapping.

Specifically, let the original low-resolution dirty image be \(\text{D}^{\text{LR}}\), and let the high-resolution reconstructed image output by the model be \({\widehat{\text{I}}}^{\text{HR}}\);\({\widehat{\text{I}}}^{\text{HR}}\) is first downsampled to the original observation scale using the scale-mapping operator \(\text{D}_{\text{s}}\text{(⋅)}\), after which the primary beam response \(\text{B}^{\text{LR}}\) and the synthesized beam \(\text{P}^{\text{LR}}\) are introduced to construct a simulated observed image, given by

\begin{equation}
{\widehat{D}}^{LR} = P^{LR} \ast \left( B^{LR} \odot D_{s}\left( {\widehat{I}}^{HR} \right) \right)
\label{eq:22}
\end{equation}

where $\odot$ denotes element-wise multiplication, and $\ast$ denotes convolution. This process is consistent with the forward degradation process in radio synthesis imaging; that is, after primary-beam modulation, PSF convolution, and scale mapping, the high-resolution reconstructed image should be able to account for the original observed low-resolution dirty image. Based on this, the observation-domain residual is defined as

\begin{equation}
R = {\widehat{D}}^{LR} - D^{LR}
\label{eq:23}
\end{equation}

where \(R\) denotes the residual distribution between the simulated observed image and the original dirty image. If the reconstruction result exhibits good observation-domain consistency, it should differ only slightly from the original dirty image after undergoing the physical degradation process described above; that is, the residual amplitudes should be low and their spatial distribution should remain relatively stable. This work uses four observation-domain consistency metrics to quantify this discrepancy: observation-domain root mean square error (Obs-RMSE), observation-domain mean absolute error (Obs-MAE), relative residual error (RRE), and residual standard deviation (\(\sigma_{\text{R}}\text{)}\).

\textbf{Observation-domain root mean square error (Obs-RMSE):} Obs-RMSE measures the overall level of squared error between the simulated observed image and the original dirty image and reflects the overall magnitude of the residual after the reconstruction result is degraded back to the observation domain. It is defined as

\begin{equation}
Obs - RMSE = \sqrt{\frac{1}{N}\sum_{i = 1}^{N}\left( I_{dirty}^{i} - {\widehat{I}}_{dirty}^{i} \right)^{2}}
\label{eq:24}
\end{equation}

where \(\text{I}_{\text{dirty}}^{\text{i}}\) denotes the intensity value of the \(i\)th pixel in the original dirty image, and \({\widehat{\text{I}}}_{\text{dirty}}^{\text{i}}\) denotes the intensity value of the corresponding pixel in the simulated observed image. A lower Obs-RMSE indicates a lower overall error between the simulated observed image and the original dirty image and, consequently, better observation-domain consistency of the reconstruction result. Because this metric is more sensitive to large errors, it can capture the effects of local anomalous responses or pronounced deviations on observation-domain consistency.

\textbf{Observation-domain mean absolute error (Obs-MAE)} \citep{article33}\textbf{:} Obs-MAE measures the mean absolute error between the simulated observed image and the original dirty image. It is less sensitive to a small number of extreme errors and is therefore better suited to reflecting the overall mean deviation. It is defined as

\begin{equation}
Obs - MAE = \frac{1}{N}\sum_{i = 1}^{N}\left| \mathbf{I}_{dirty}^{i} - {\widehat{\mathbf{I}}}_{dirty}^{i} \right|
\label{eq:25}
\end{equation}

A lower Obs-MAE indicates a lower mean residual between the simulated observed image and the original dirty image. This metric can be used to assess whether the reconstruction result remains close overall to the original observational data after being degraded back to the observation domain, while avoiding the disproportionate influence of a small number of pixels with large errors that can arise when only squared error is used.

\textbf{Relative residual error (RRE)} \citep{article34}\textbf{:} RRE measures the ratio of the observation-domain residual to the intensity of the original observed signal. Because it is normalized by the overall image intensity, it is more suitable for comparing observation-domain consistency across different samples, intensity ranges, or super-resolution factors. It is defined as

\begin{equation}
RRE = \frac{\left\| \mathbf{I}_{dirty} - {\widehat{\mathbf{I}}}_{dirty} \right\|_{2}}{\left\| \mathbf{I}_{dirty} \right\|_{2} + \epsilon}
\label{eq:26}
\end{equation}

Where \(\| \cdot \|_{2}\) denotes the \(\text{L}_{\text{2}}\) norm, and \(\epsilon\) is a small constant used to prevent the denominator from becoming zero. A lower RRE indicates a smaller relative deviation of the simulated observed image from the original dirty image, showing that the high-resolution structures generated by the model can better account for the original observational data after being reprojected into the observation domain.

\textbf{Observation-domain residual standard deviation (}\(\sigma_{\text{R}}\)\textbf{)} \citep{article35}\textbf{:} The residual standard deviation quantifies the dispersion of the residual distribution between the simulated observed image and the original dirty image. Let the observation-domain residual be given by

\begin{equation}
R = I_{dirty} - {\widehat{I}}_{dirty}
\label{eq:27}
\end{equation}

The residual standard deviation is then defined as

\begin{equation}
\sigma_{R} = Std(R)
\label{eq:28}
\end{equation}

where \(\text{Std}\text{(⋅)}\) denotes the standard deviation operation. A lower residual standard deviation indicates a more concentrated observation-domain residual distribution and a more stable error between the simulated observed image and the original dirty image. Conversely, a larger residual standard deviation indicates more pronounced spatial fluctuations in the residual, which may suggest problems such as local spurious responses, structural shifts, or noise amplification.

In summary, the four metrics described above are all error-based metrics: Obs-RMSE primarily reflects the overall squared error and is more sensitive to large residuals; Obs-MAE reflects the mean absolute deviation and is relatively robust to extreme errors; RRE measures the ratio of the residual to the original observed signal, facilitating comparisons across samples and super-resolution factors; and \(\sigma_{\text{R}}\) evaluates the stability of the residual distribution. Together, these four metrics provide a comprehensive assessment of whether the reconstruction result can adequately account for the original observational data from the four perspectives of overall error, mean deviation, relative error, and residual variability. It should be noted that observation-domain consistency metrics primarily evaluate the physical consistency of the reconstructed image after it is degraded back to the observation domain and cannot completely replace image-domain metrics in assessing the structural integrity and detail-recovery quality of high-resolution structures. Therefore, this work combines image-domain metrics with observation-domain consistency metrics to provide a more comprehensive evaluation of the model's reconstruction performance and physical reliability.
\subsection{Ablation Experiments}\label{sec:ablation}

To verify the effectiveness and necessity of each module in PhySR for super-resolution reconstruction, we conducted ablation experiments. The ablation experiments focused on three core modules---the physical consistency constraint, the dynamic cascaded upsampling module, and the multiscale feature residual module---by removing each module in turn from the full model and comparing the resulting variants with the full PhySR model, while keeping the training data, number of training epochs, optimization strategy, and testing procedure unchanged. The experiments were evaluated from both quantitative and qualitative perspectives. Quantitatively, image-domain metrics including PSNR, SSIM, and RMSE, together with observation-domain consistency metrics including Obs-RMSE, Obs-MAE, RRE, and \(\sigma_{\text{R}}\), were used to compare the changes in reconstruction accuracy, error suppression, and physical consistency among the different model variants. Qualitatively, comparative images of the reconstruction results were used to further analyze the effects of the different modules on the recovery of extended-source contours, small-scale structures, and mixed structures, as well as on effect elimination. Through these comparisons, the contribution of each module to the overall performance of PhySR can be assessed more comprehensively.

This study compared four model configurations: (1) the full PhySR model; and (2) the w/o PINN model, in which the physical consistency constraint was removed, the dynamic cascaded upsampling module and the multiscale feature residual module were retained, and model training relied solely on image-domain loss constraints. This experiment was designed to verify the role of the physical constraints in suppressing spurious structures, improving observation-domain consistency, and enhancing reconstruction stability. (3) In the w/o Cascaded Upsampling model, the dynamic cascaded upsampling module was removed, the PINN and the multiscale feature residual module were retained, and the progressive upsampling process was replaced with a conventional upsampling scheme. This experiment was designed to verify the role of the dynamic cascaded upsampling module in cross-scale feature transfer, progressive resolution enhancement, and detail recovery. (4) In the w/o Multi-scale Residual model, the multiscale feature residual module was removed, while the PINN and the dynamic cascaded upsampling module were retained. This experiment was designed to verify the contribution of the multiscale residual structure to the recovery of small-scale structures, edge contours, and local details.

To verify the effectiveness of the physical consistency constraint, the dynamic cascaded upsampling module, and the multiscale feature residual module in PhySR, we comparatively analyzed the reconstruction results of the full model and the three ablated models at different super-resolution factors. Figure~\ref{fig:ablation} presents the reconstruction results of the four models for extended-source and mixed-source structures, with panels (a), (b), and (c) corresponding to the 2×, 4×, and 8× super-resolution tasks, respectively, to assess the effects of each module on structural fidelity, detail recovery, and artifact suppression under different degradation levels. The full PhySR model maintains good reconstruction performance in the 2×, 4×, and 8× super-resolution tasks. For extended sources, PhySR fully recovers the main morphology while preserving continuous structural contours and clear edge transitions; for mixed sources, PhySR accurately recovers the central extended source while also clearly reconstructing the surrounding fine structures at small scales, demonstrating good small-scale structure recovery capability.

\insertwidefigure{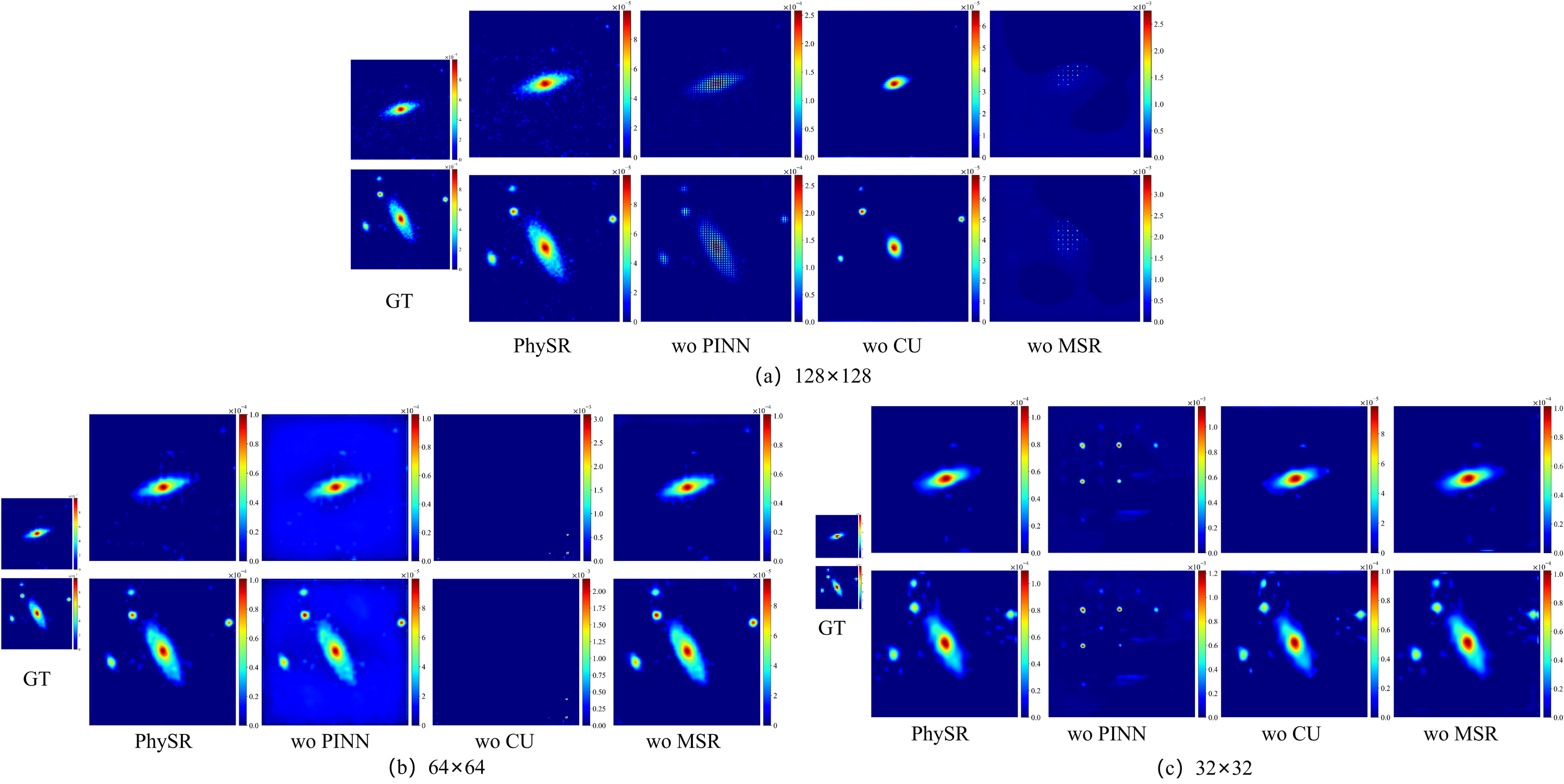}{0.94\textwidth}{6.5cm}{Comparison of the ablation experiment results. Panels (a), (b), and (c) show the results for the 2×, 4×, and 8× super-resolution tasks, respectively. Each group of results presents two typical astronomical source structures: the first row shows an extended-source structure, and the second row shows a mixed-source structure. In panel (a), the clean image has a resolution of 128×128 pixels, and the reconstructed image has a resolution of 256×256 pixels. In panel (b), the clean image has a resolution of 64×64 pixels, and the reconstructed image has a resolution of 256×256 pixels. In panel (c), the clean image has a resolution of 32×32 pixels, and the reconstructed image has a resolution of 256×256 pixels.}{fig:ablation}

By contrast, in the 2× super-resolution task, the w/o PINN model without the physical consistency constraint produces a discontinuous internal brightness distribution in the extended source and blurred recovery of small-scale local details in the mixed source, while the reconstructed structures of both source types exhibit pronounced grid-like dotted textures and regular stripe artifacts. In the 4× super-resolution task, pronounced spurious responses are present in the backgrounds of the reconstructed extended and mixed sources. In the 8× super-resolution task, the reconstruction completely loses the source structures and details. These results indicate that the physical consistency constraint effectively suppresses spurious textures and background artifacts, preserves the continuity of source structures, and improves the model's reconstruction stability under different degradation levels. In the 2× super-resolution task, the w/o\_CU model without the dynamic cascaded upsampling module incompletely recovers the central structures of both the extended and mixed sources; the small-scale structures in the mixed source are blurred, and local details are missing. In the 4× super-resolution task, the reconstruction completely fails to recover valid source-structure information. In the 8× super-resolution task, although the model recovers the approximate contours of the extended and mixed sources, the small-scale local details remain incomplete, and pronounced ring-like artifacts appear along the image edges. These results verify that the dynamic cascaded upsampling module effectively enhances the progressive transfer of cross-scale features, enabling the model to better recover source-structure contours and local details during resolution enhancement and improving the stability and robustness of high-factor super-resolution reconstruction. In the 2× super-resolution task, the w/o\_MSR model without the multiscale feature residual module is almost unable to recover the source structures; in the 4× and 8× tasks, although it approximately recovers the main structures and small-scale details of the extended and mixed sources, pronounced artifacts appear along the image edges. This indicates that the multiscale feature residual module helps enhance local-detail representation, enabling the model to better recover small-scale structures and reduce edge artifacts.

This work uses peak signal-to-noise ratio (PSNR), structural similarity index measure (SSIM), and root mean square error (RMSE) to evaluate image reconstruction performance in the image domain (Table~\ref{tab:ablation_image}). Across the 2×, 4×, and 8× super-resolution tasks, the full PhySR model achieves the best values for PSNR, SSIM, and RMSE, whereas all three ablated models exhibit marked deterioration in these metrics. For example, in the 4× super-resolution task, the full PhySR model achieves a PSNR of 44.65 dB, an SSIM of 0.9940, and an RMSE of 0.0065; compared with the w/o PINN model, PSNR and SSIM increase by 24.82 dB and 0.7605, respectively, while RMSE decreases by 0.0958. By contrast, the w/o\_CU model achieves a PSNR of only 19.84 dB, while its SSIM decreases to 0.5751 and its RMSE increases to 0.1050. The w/o\_MSR model also exhibits a certain degree of performance degradation across all three metrics. These results demonstrate that the physical consistency constraint, the dynamic cascaded upsampling module, and the multiscale feature residual module all play important roles in the reconstruction performance of PhySR.

\begin{table*}[!tb]
\centering

\begin{minipage}[t]{0.49\textwidth}
\vspace{0pt}

\captionof{table}{Comparison of image-domain metrics for the ablation experiments.}
\label{tab:ablation_image}

\centering

{\fontsize{7.0}{7.8}\selectfont
\setlength{\tabcolsep}{1.0pt}
\renewcommand{\arraystretch}{1.04}

\begin{tabular*}{0.98\linewidth}
{@{\extracolsep{\fill}}llcccccc@{}}
\hline\hline
Scale & Model & PINN & CU & MSR & PSNR & SSIM & RMSE \\
\hline

\multirow{4}{*}{$\times2$}
 & w/o PINN
 & $\times$
 & $\checkmark$
 & $\checkmark$
 & 37.23dB
 & 0.8903
 & 0.0351 \\

 & w/o CU
 & $\checkmark$
 & $\times$
 & $\checkmark$
 & 24.69dB
 & 0.7164
 & 0.0607 \\

 & w/o MSR
 & $\checkmark$
 & $\checkmark$
 & $\times$
 & 20.17dB
 & 0.5388
 & 0.1041 \\

 & \textbf{PhySR}
 & $\checkmark$
 & $\checkmark$
 & $\checkmark$
 & \textbf{45.20dB}
 & \textbf{0.9945}
 & \textbf{0.0059} \\
\hline

\multirow{4}{*}{$\times4$}
 & w/o PINN
 & $\times$
 & $\checkmark$
 & $\checkmark$
 & 19.83dB
 & 0.2335
 & 0.1023 \\

 & w/o CU
 & $\checkmark$
 & $\times$
 & $\checkmark$
 & 19.84dB
 & 0.5751
 & 0.1050 \\

 & w/o MSR
 & $\checkmark$
 & $\checkmark$
 & $\times$
 & 36.07dB
 & 0.9466
 & 0.0165 \\

 & \textbf{PhySR}
 & $\checkmark$
 & $\checkmark$
 & $\checkmark$
 & \textbf{44.65dB}
 & \textbf{0.9940}
 & \textbf{0.0065} \\
\hline

\multirow{4}{*}{$\times8$}
 & w/o PINN
 & $\times$
 & $\checkmark$
 & $\checkmark$
 & 18.07dB
 & 0.1124
 & 0.1267 \\

 & w/o CU
 & $\checkmark$
 & $\times$
 & $\checkmark$
 & 38.04dB
 & 0.9682
 & 0.0159 \\

 & w/o MSR
 & $\checkmark$
 & $\checkmark$
 & $\times$
 & 33.79dB
 & 0.8781
 & 0.0321 \\

 & \textbf{PhySR}
 & $\checkmark$
 & $\checkmark$
 & $\checkmark$
 & \textbf{39.19dB}
 & \textbf{0.9700}
 & \textbf{0.0138} \\
\hline
\end{tabular*}

}
\end{minipage}
\hfill
\begin{minipage}[t]{0.49\textwidth}
\vspace{0pt}

\captionof{table}{Comparison of observation-domain metrics for the ablation experiments.}
\label{tab:ablation_observation}

\centering

{\fontsize{7.0}{7.8}\selectfont
\setlength{\tabcolsep}{1.0pt}
\renewcommand{\arraystretch}{1.04}

\begin{tabular*}{0.98\linewidth}
{@{\extracolsep{\fill}}llcccc@{}}
\hline\hline
Scale & Model & Obs\_RMSE & Obs\_MAE & RRE & $\sigma_R$ \\
\hline

\multirow{4}{*}{$\times2$}
 & w/o PINN
 & 5.62e-06
 & 3.09e-06
 & 7.40e-01
 & 5.39e-06 \\

 & w/o CU
 & 5.43e-06
 & 3.11e-06
 & 7.37e-01
 & 5.17e-06 \\

 & w/o MSR
 & 1.97e-05
 & 1.18e-05
 & 2.66e-00
 & 1.97e-05 \\

 & \textbf{PhySR}
 & \textbf{4.31e-06}
 & \textbf{2.94e-06}
 & \textbf{5.80e-01}
 & \textbf{4.02e-06} \\
\hline

\multirow{4}{*}{$\times4$}
 & w/o PINN
 & 4.81e-06
 & 3.23e-06
 & 4.52e-01
 & 4.42e-06 \\

 & w/o CU
 & 1.19e-05
 & 5.71e-06
 & 1.14e+00
 & 1.17e-05 \\

 & w/o MSR
 & 4.53e-06
 & 3.27e-06
 & 4.21e-01
 & 4.07e-06 \\

 & \textbf{PhySR}
 & \textbf{4.48e-06}
 & \textbf{3.15e-06}
 & \textbf{4.16e-01}
 & \textbf{4.02e-06} \\
\hline

\multirow{4}{*}{$\times8$}
 & w/o PINN
 & 2.10e-05
 & 9.70e-06
 & 9.18e-01
 & 2.09e-05 \\

 & w/o CU
 & 7.26e-06
 & 4.77e-06
 & 6.28e-01
 & 7.17e-06 \\

 & w/o MSR
 & 6.87e-06
 & 4.62e-06
 & 6.05e-01
 & 6.74e-06 \\

 & \textbf{PhySR}
 & \textbf{6.56e-06}
 & \textbf{4.51e-06}
 & \textbf{5.93e-01}
 & \textbf{6.52e-06} \\
\hline
\end{tabular*}

}
\end{minipage}

\end{table*}

Building on the image-domain metric evaluation, this work further evaluates the effectiveness of the modules from the perspective of observation-domain consistency (Table~\ref{tab:ablation_observation}). The results show that the joint use of the physical consistency constraint, the dynamic cascaded upsampling module, and the multiscale feature residual module effectively reduces the residuals obtained after the reconstruction results are degraded back to the observation domain, makes the reconstructed images more consistent with the physical observation process, and thereby improves the model's observation-domain consistency and physical reliability. Specifically, the full PhySR model achieves the best values for all four metrics in the 2×, 4×, and 8× super-resolution tasks. More specifically, compared with w/o PINN, the full PhySR model substantially reduces the observation-domain residuals at all three super-resolution factors; particularly in the 8× task, Obs-RMSE decreases from 2.10e-05 to 6.56e-06, Obs-MAE from 9.70e-06 to 4.51e-06, RRE from 9.18e-01 to 5.93e-01, and \(\sigma_{\text{R}}\) from 2.09e-05 to 6.52e-06. Compared with w/o CU, the full PhySR model yields markedly lower values for all four metrics in the 4× task. Compared with w/o MSR, the full PhySR model shows the most pronounced advantage in the 2× task, with Obs-RMSE, Obs-MAE, RRE, and \(\sigma_{\text{R}}\) decreasing from 1.97e-05, 1.18e-05, 2.66e+00, and 1.97e-05 to 4.31e-06, 2.94e-06, 5.80e-01, and 4.02e-06, respectively.

\FloatBarrier

\subsection{Experimental Results and Analysis}\label{sec:analysis}

To ensure the objectivity and fairness of the experimental results, all tests were conducted under the same experimental environment and using identical data splits. To comprehensively verify the effectiveness of the proposed PhySR model, we compared it with existing general-purpose methods and deep learning methods for super-resolution and analyzed the experimental results in terms of both quantitative metrics and reconstruction quality. This work focuses on the 4× and 8× super-resolution tasks, while the results for the 2× task are used only as supplementary evidence for overall performance validation, to evaluate the model's capabilities in structural recovery, detail reconstruction, and the elimination of coupling effects under strong- and extreme-degradation conditions, respectively. Specifically, we first present the reconstruction results of PhySR for the 4× and 8× tasks, then compare them with those obtained using existing general-purpose methods that combine CLEAN-type algorithms with primary beam correction and interpolation-based upsampling, and finally conduct quantitative and qualitative comparisons with mainstream deep learning methods for super-resolution, thereby validating the effectiveness and advantages of the proposed method from multiple perspectives.

\begin{figure}[!tbp]
    \centering
    \includegraphics[
        width=0.98\columnwidth,
        keepaspectratio
    ]{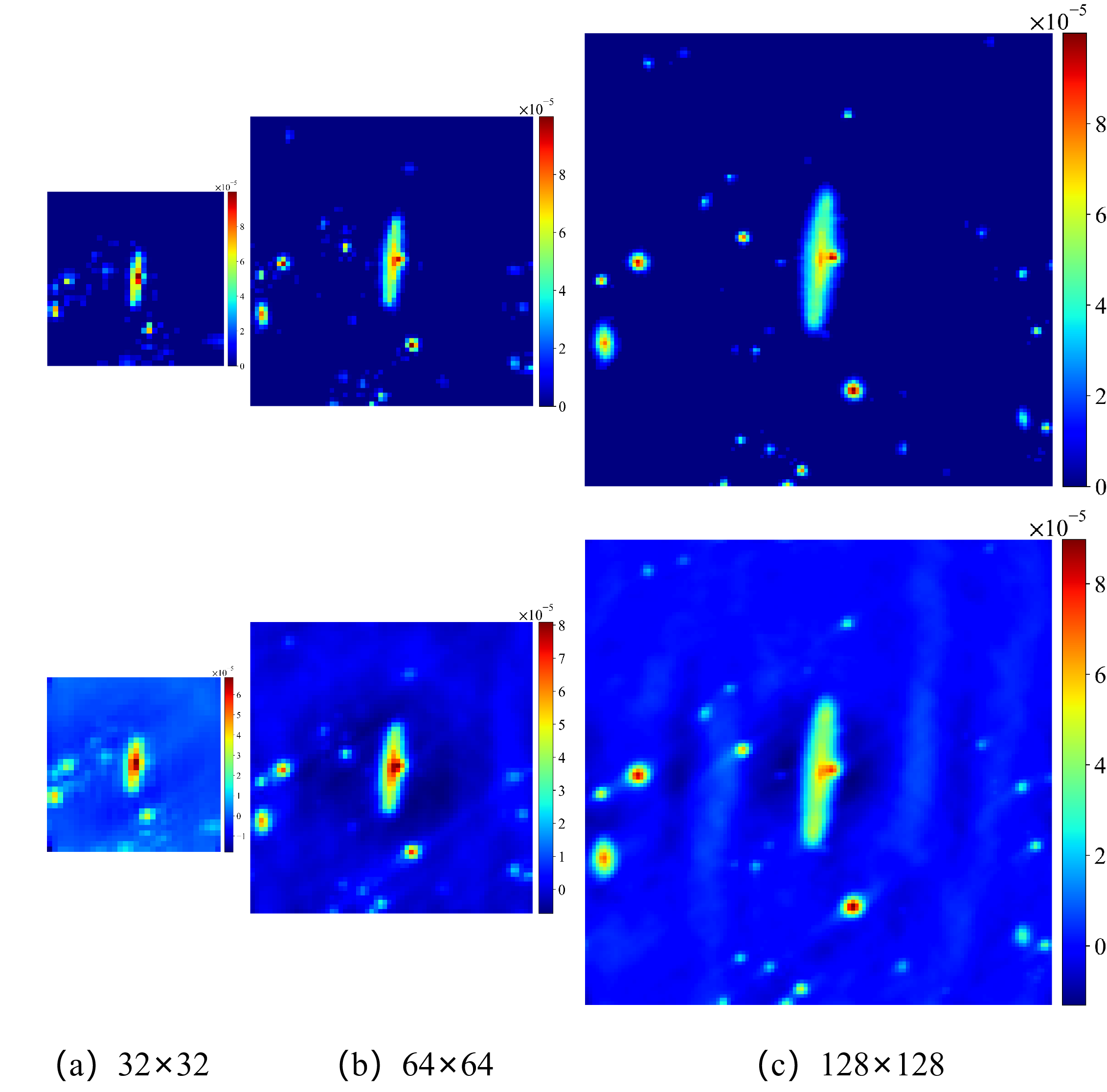}
    \caption{Simulated SKA-Mid observations: (a) clean and dirty images at a resolution of 32×32 pixels; (b) clean and dirty images at a resolution of 64×64 pixels; and (c) clean and dirty images at a resolution of 128×128 pixels.}
    \label{fig:inputs}
\end{figure}

Different input resolutions correspond to degradation scenarios of different severities, and changes in resolution directly affect the ability of an image to represent structures and preserve details, while the coupling effects between the primary and synthesized beams further suppress small-scale target details, weaken weak-source responses, and intensify artifacts and background contamination. To analyze more thoroughly the quality characteristics of the input images and their effects on the subsequent reconstruction results, we examine the image characteristics under different degradation scenarios (Fig.~\ref{fig:inputs}). Under the extreme-degradation scenario shown in Fig.~\ref{fig:inputs}(a), the clean image retains only the basic outline of the main target structure, whereas the dirty image exhibits relatively coarse edge transitions, poorly resolved levels of local detail, and surrounding weak-source information that is difficult to distinguish effectively. In Fig.~\ref{fig:inputs}(b), as the resolution increases, the overall blurring of the clean image is reduced, ring-like artifacts resembling concentric circles appear around the central extended source in the dirty image, and the pixel intensities of its point sources are attenuated. A comparison of Figs.~\ref{fig:inputs}(c) and \ref{fig:inputs}(b) shows that, with a further increase in resolution, the overall structure and surrounding point-source information in the clean image in Fig.~\ref{fig:inputs}(c) are more fully preserved, and the main morphology and brightness distribution become clearer; meanwhile, the ring-like artifacts around the central extended source in the dirty image become more continuous, and the attenuation of the pixel intensities of the surrounding point sources becomes more pronounced.

\subsubsection{Analysis of PhySR Reconstruction Results under Different Degradation Scenarios}\label{sec:physr_results}

\begin{figure}[!tbp]
    \centering
    \includegraphics[
        width=0.98\columnwidth,
        keepaspectratio
    ]{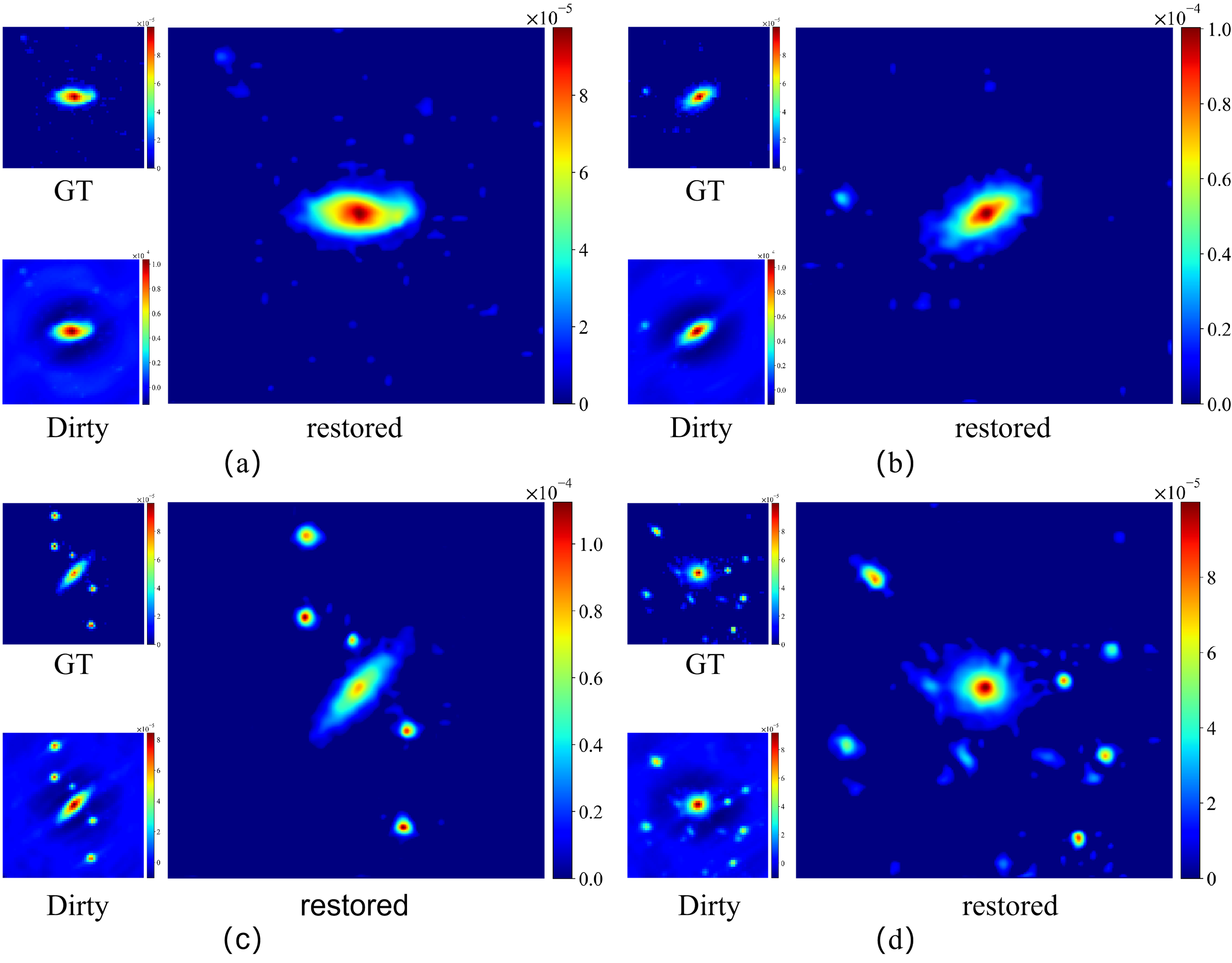}
    \caption{Comparison of PhySR 4× super-resolution reconstruction results for real sky sources. Panels (a) and (b) show extended-source structures, whereas panels (c) and (d) show mixed-source structures. The clean and dirty images have a resolution of 64×64 pixels, whereas the reconstructed images have a resolution of 256×256 pixels.}
    \label{fig:physr4x}
\end{figure}

PhySR exhibits good reconstruction stability in the strong-degradation scenario; for both extended-source and mixed-source structures, the model effectively removes pronounced artifacts and suppresses spurious background responses, fully recovers the main morphology of the central source, preserves clear edge contours, and accurately reconstructs small-scale structures and local details, demonstrating strong structural-fidelity and detail-reconstruction capabilities (Fig.~\ref{fig:physr4x}). Figs.~\ref{fig:physr4x}(a) and \ref{fig:physr4x}(b) show the reconstruction results for extended sources: owing to the limited resolution, the central extended source in the clean image (GT) is blurred to some extent, whereas under the coupling effects, the main structure in the dirty image becomes more severely blurred and broadened, ring-like artifacts appear around the central source, and pronounced spurious responses emerge in the background. By contrast, the image reconstructed by PhySR not only accurately removes the ring-like artifacts and noise introduced by the coupling effects, but also fully restores the overall shape of the central extended source, with smoother and more continuous edge transitions and a clearer overall structure. Figs.~\ref{fig:physr4x}(c) and \ref{fig:physr4x}(d) show the reconstruction results for mixed sources: owing to the resolution limitation and coupling effects, the small-scale structures and local details in the clean image are blurred, while some weak sources and small-scale structures in the dirty image are obscured by artifacts and background noise. PhySR, however, effectively reconstructs the main shape of the central source, suppresses spurious background responses, and accurately recovers the surrounding weak-source structures and fine small-scale details, demonstrating good small-scale structure reconstruction capability.

To facilitate comparison of structural recovery in the 8× super-resolution task, all clean and dirty images for this task are displayed at the same scale as the reconstructed images, together with zoomed-in views for analyzing differences in detail. It should be noted that this processing is used solely for visualization and does not alter the original image resolution.

\begin{figure}[!t]
\centering
\includegraphics[
    width=0.98\columnwidth,
    keepaspectratio
]{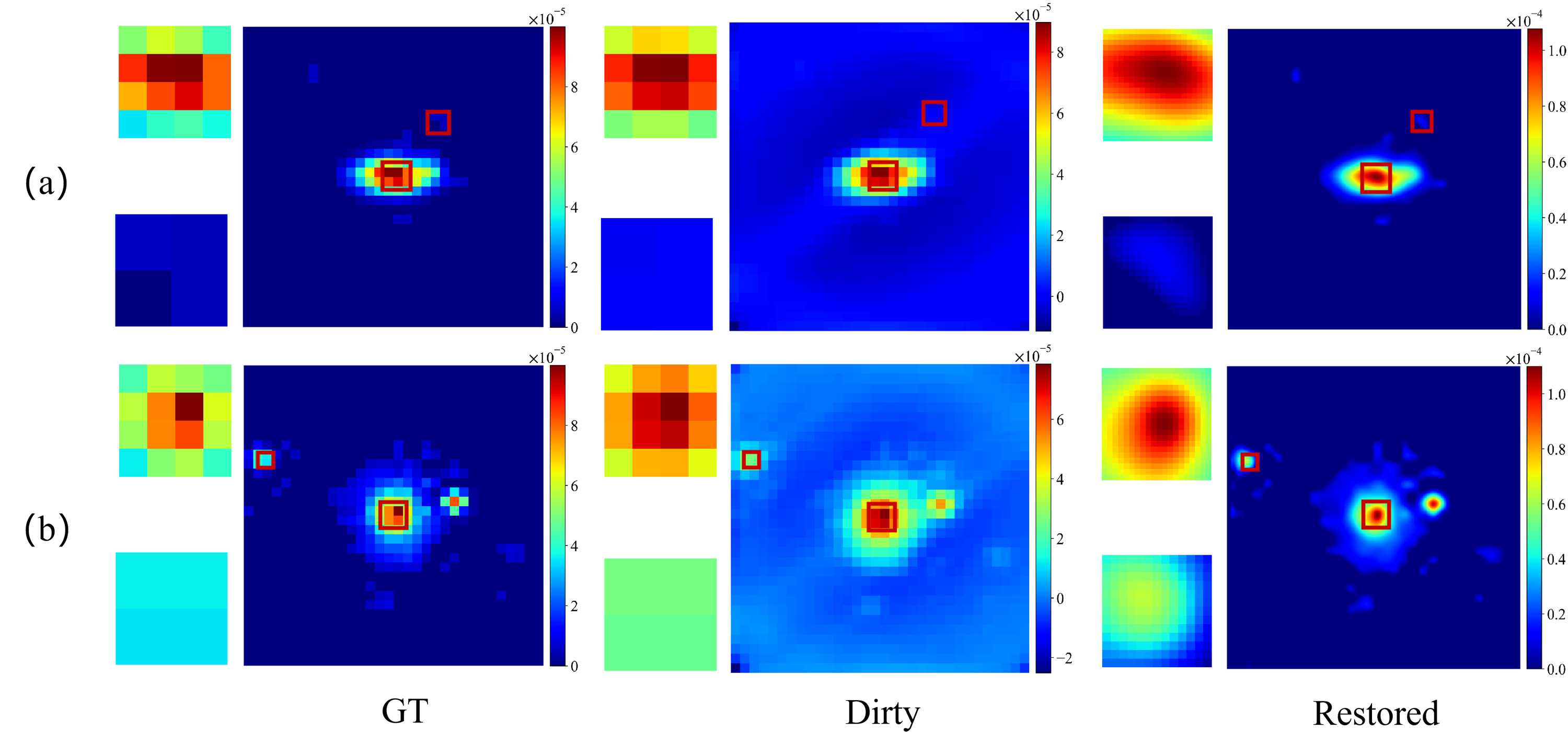}
\caption{Comparison of PhySR $8\times$ super-resolution reconstruction
results for real sky sources. Panel (a) shows an extended-source structure,
whereas panel (b) shows a mixed-source structure. The clean and dirty images
have a resolution of $32\times32$ pixels, whereas the reconstructed images
have a resolution of $256\times256$ pixels. The red boxes indicate the
large-scale morphology of the central source (large box) and the fine
small-scale details (small box), respectively.}
\label{fig:physr8x}
\end{figure}

The proposed method maintains relatively stable reconstruction performance even under the extreme-degradation scenario; it not only recovers the main structures of extended sources but also effectively reconstructs the small-scale details in mixed sources, demonstrating strong capabilities in structural preservation and small-scale detail reconstruction (Fig.~\ref{fig:physr8x}). As can be seen from the extended-source case in Fig.~\ref{fig:physr8x}(a), the central extended source is severely blurred in both the clean and dirty images, with local details being almost invisible; moreover, owing to the limited resolution, the central extended source and local details exhibit pronounced pixelation overall. After reconstruction by PhySR, the main morphology of the central extended source is recovered relatively completely, the edge transitions become more continuous, the pixelation caused by the limited resolution disappears completely, the recovered local details are smooth, and the overall resolution is greatly improved. In the mixed-source case shown in Fig.~\ref{fig:physr8x}(b), the central source and fine small-scale details likewise exhibit pronounced pixelation; only the approximate structure of the central source is barely discernible, while the structural shapes of the small-scale details are completely lost under the coupling effects and resolution limitation. By contrast, in the image reconstructed by PhySR, the structure of the central source is fully recovered, the structure at the source center is continuous and clear, and the transitions between regions of different intensities are smooth. At the same time, even when the small-scale details are completely indistinguishable, the model fully reconstructs their overall structures, with relatively smooth edge contours, largely eliminates the pixelation, and likewise greatly improves the overall resolution.


\subsubsection{Comparative Analysis with Existing General-Purpose Methods}
\label{sec:general_comparison}

This work selects three representative CLEAN algorithms---Högbom CLEAN, MS-CLEAN, and ASP-CLEAN---in combination with image-domain primary beam correction (PBC) as general-purpose methods for eliminating the coupling effects between the primary and synthesized beams. Image-domain primary beam correction performs pixel-by-pixel gain compensation on the image after imaging using a preconstructed primary beam model, thereby mitigating the brightness attenuation and spatial distortion caused by the primary-beam effect \citep{article36}. Meanwhile, bilinear interpolation upsampling (BIU) \citep{article37} is adopted as a general-purpose method for increasing the resolution; as one of the classical resampling methods for image scaling, bilinear interpolation is simple to implement and produces relatively smooth interpolation results and is therefore widely used in image upsampling and scale-transformation tasks. It should be noted that all existing general-purpose methods adopt a three-step processing procedure: a CLEAN algorithm is first used to eliminate the synthesized-beam effect, image-domain primary beam correction is then applied to mitigate the primary-beam effect, and BIU is subsequently used to upsample the reconstructed image to the corresponding resolution.

\insertwidefigure{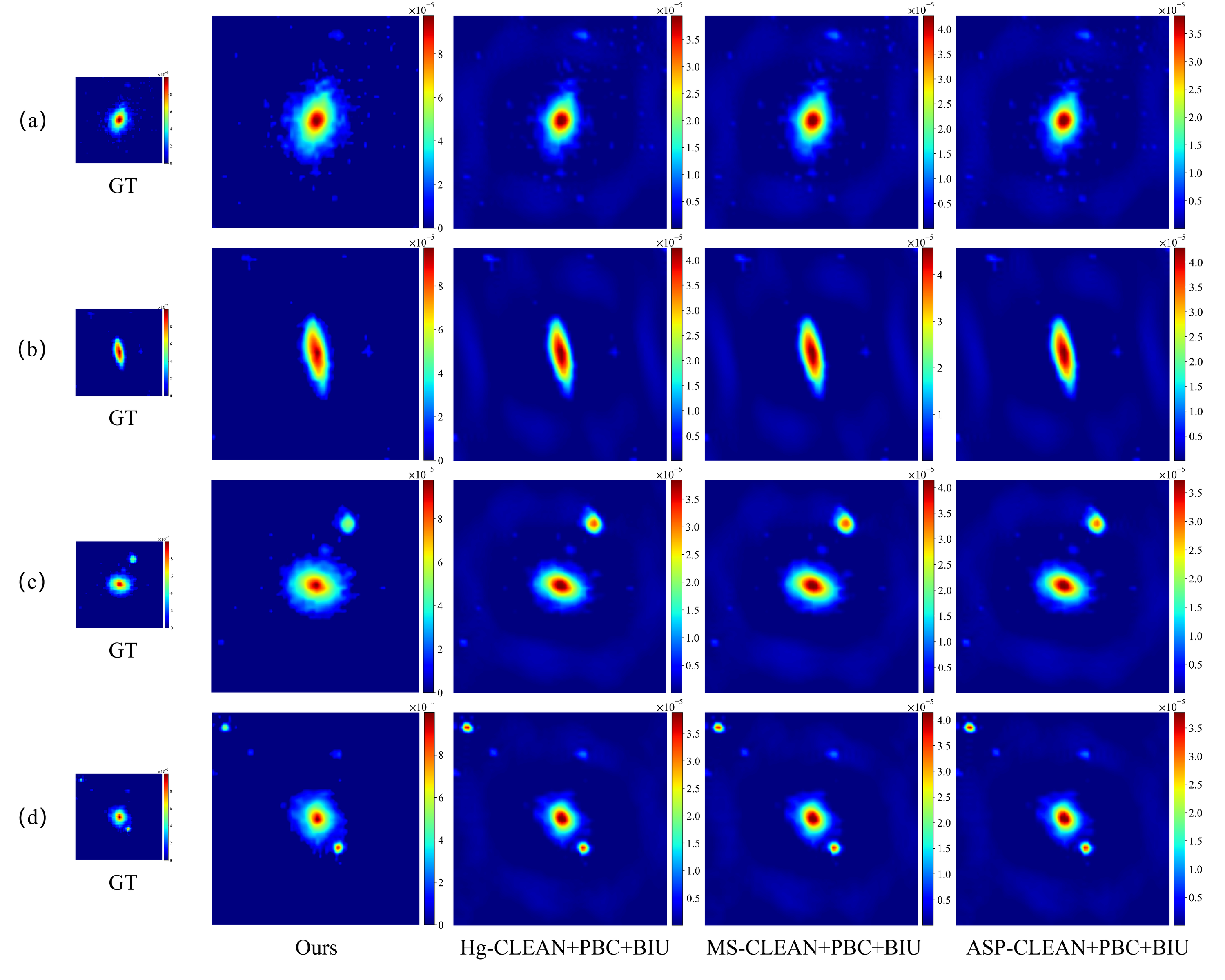}{0.94\textwidth}{7.5cm}{Comparison of 4× super-resolution reconstruction results obtained with PhySR and existing general-purpose methods. Panels (a) and (b) show extended-source structures, whereas panels (c) and (d) show mixed-source structures. The clean images have a resolution of 64×64 pixels, whereas the reconstructed images produced by the different methods have a resolution of 256×256 pixels.}{fig:general4x}

Under the strong-degradation scenario, although existing general-purpose methods can achieve basic structural recovery and image upscaling, their capabilities for recovering edge structures and small-scale details and suppressing artifacts under complex degradation conditions remain limited, resulting in incomplete recovery of extended-source edges, insufficient reconstruction of small-scale structures, and residual ring-like artifacts in the reconstructed images. By contrast, PhySR more effectively suppresses the coupling effects between the primary and synthesized beams, preserves the integrity of source structures, and recovers weak sources and small-scale details, thereby verifying its effectiveness under complex strong-degradation conditions (Fig.~\ref{fig:general4x}). Figs.~\ref{fig:general4x}(a) and \ref{fig:general4x}(b) show the reconstruction results for extended sources; although the CLEAN+PBC+BIU methods can still approximately recover the main structure of the central extended source, its edge contours are not fully recovered, and ring-like artifacts caused by the coupling effects remain around the source. By contrast, the results reconstructed by PhySR not only fully recover the main structure of the central extended source, with clear and smooth edge transitions, but also eliminate the effects of the coupling between the primary and synthesized beams. Figs.~\ref{fig:general4x}(c) and \ref{fig:general4x}(d) show the reconstruction results for mixed sources; the surrounding weak sources and small-scale structures are not adequately recovered by the existing general-purpose methods, local details are more readily lost after interpolation-based upsampling, and spurious responses and artifacts of varying severity remain in the background regions. In comparison, the PhySR reconstruction results accurately recover the small-scale details and weak-source structures and eliminate the spurious background responses and artifacts.

In the 8× super-resolution task, we compare the zoomed-in reconstruction results obtained with PhySR with those obtained using the existing general-purpose methods (Fig.~\ref{fig:general8x}), with panel (a) showing an extended-source case and panel (b) showing a mixed-source case. As can be seen from the figure, PhySR clearly outperforms the existing general-purpose methods in reconstructing both extended and mixed sources, with the morphology and structure of the central source clearly recovered and the small-scale details restored relatively completely. By contrast, although the three existing general-purpose methods recover the approximate structure of the extended source, the zoomed-in views reveal residual pixelation at the source center, insufficiently continuous edges of the high-intensity regions, signal loss in the small-scale structures of the mixed source, incomplete structural recovery, and a certain degree of residual artifacts along the image edges. Overall, PhySR maintains good capabilities for structural preservation and detail recovery in the 8× super-resolution task.

\insertwidefigure
{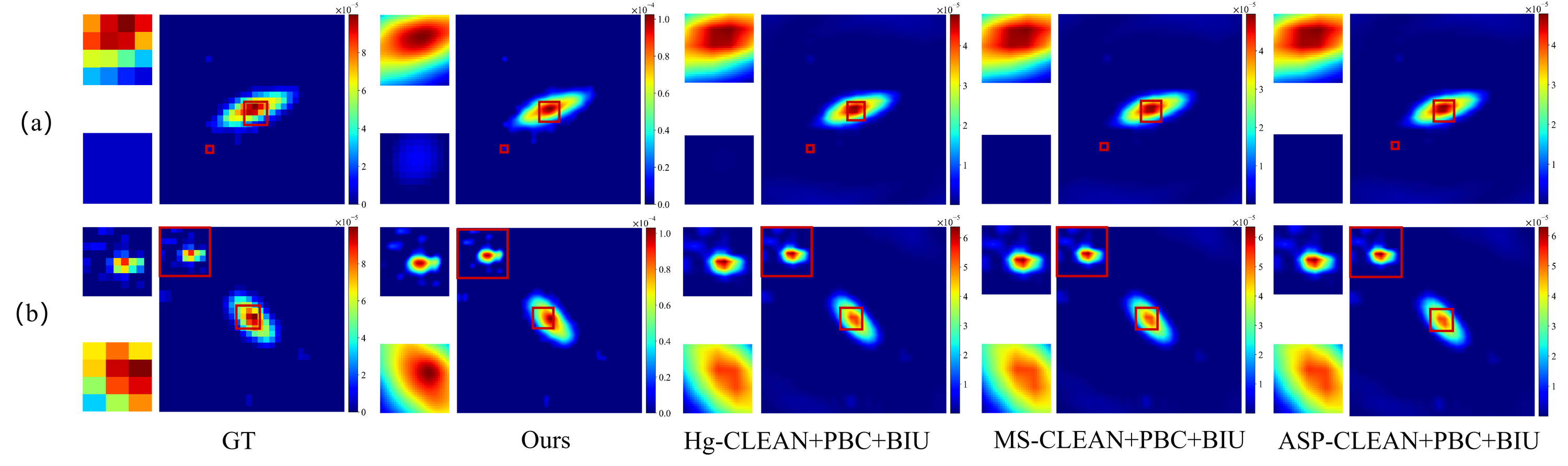}
{0.94\textwidth}
{5.0cm}
{Comparison of 8× super-resolution reconstruction results obtained with PhySR and existing general-purpose methods. Panel (a) shows an extended-source structure, whereas panel (b) shows a mixed-source structure. The clean images have a resolution of 32×32 pixels, whereas the reconstructed images produced by the different methods have a resolution of 256×256 pixels. The red boxes indicate the large-scale structure of the central source and the small-scale details, respectively.}
{fig:general8x}

\begin{table}[!tbp]
\caption{Comparison of image-domain metrics between PhySR and existing general-purpose methods.}
\label{tab:general_methods}
\centering
\scriptsize
\setlength{\tabcolsep}{2.2pt}
\renewcommand{\arraystretch}{1.08}

\resizebox{\columnwidth}{!}{%
\begin{tabular}{llccc}
\hline\hline
Scale & Model & PSNR & SSIM & RMSE \\
\hline
\multirow{4}{*}{$\times4$}
 & Hg-CLEAN+PBC+BIU  & 31.37 & 0.6238 & 0.0275 \\
 & MS-CLEAN+PBC+BIU  & 31.77 & 0.6137 & 0.0262 \\
 & ASP-CLEAN+PBC+BIU & 31.13 & 0.6164 & 0.0284 \\
 & \textbf{PhySR}     & \textbf{44.65} & \textbf{0.9940} & \textbf{0.0065} \\
\hline
\multirow{4}{*}{$\times8$}
 & Hg-CLEAN+PBC+BIU  & 34.68 & 0.8640 & 0.0188 \\
 & MS-CLEAN+PBC+BIU  & 35.30 & 0.8724 & 0.0175 \\
 & ASP-CLEAN+PBC+BIU & 34.72 & 0.8648 & 0.0187 \\
 & \textbf{PhySR}     & \textbf{39.19} & \textbf{0.9700} & \textbf{0.0138} \\
\hline
\end{tabular}%
}
\end{table}

The comparison of the mean image-domain metric values of PhySR and the existing general-purpose methods in the 4× and 8× super-resolution tasks further verifies the reconstruction advantages of the proposed method under different degradation scenarios, with PhySR outperforming the existing general-purpose methods in image fidelity, structural preservation, and error control. At the same time, the quantitative results also show that the existing general-purpose methods are constrained by their stepwise processing procedure and still suffer from error accumulation and insufficient detail recovery under complex degradation conditions. Specifically, in the 4× super-resolution task, PhySR achieves the best results among all methods for the three performance metrics---PSNR, SSIM, and RMSE---with values of 44.65 dB, 0.9940, and 0.0065, respectively, outperforming Hg-CLEAN+PBC+BIU, which obtains 31.37 dB, 0.6238, and 0.0275; MS-CLEAN+PBC+BIU, which obtains 31.77 dB, 0.6137, and 0.0262; and ASP-CLEAN+PBC+BIU, which obtains 31.13 dB, 0.6164, and 0.0284. This demonstrates that PhySR retains strong reconstruction stability and fidelity under the strong-degradation scenario. In the 8× super-resolution task, PhySR continues to outperform the other methods in PSNR, SSIM, and RMSE; compared with MS-CLEAN+PBC+BIU, the best-performing traditional general-purpose method, PhySR improves PSNR by 6.74 dB and SSIM by 0.1044 while reducing RMSE by 0.0052.

\FloatBarrier

\subsubsection{Comparative Analysis with Mainstream Deep Learning Methods for Super-Resolution}
\label{sec:dl_comparison}

This work selects five mainstream deep learning models for super-resolution for comparison with PhySR. Among them, SwinIR \citep{article38} is built around residual Swin Transformer blocks and effectively improves feature representation and reconstruction quality in image super-resolution by combining local-window attention with the shifted-window mechanism. Building on SwinIR, HAT \citep{article39} further integrates channel attention with window-based self-attention and introduces an overlapping cross-attention mechanism to enhance cross-window information interaction and detail recovery. SRFormer \citep{article40} introduces a permuted self-attention mechanism to enlarge the attention window while controlling computational complexity, thereby balancing reconstruction performance and computational efficiency. To address the information bottleneck in deep networks, DRCT \citep{article41} incorporates dense residual connections within residual groups to improve the stability of deep-feature propagation and the ability to preserve spatial information. SR3 \citep{article42} introduces diffusion models into super-resolution and progressively generates high-resolution results through conditional iterative denoising, offering strong advantages in perceptual quality and detail generation.

In comparison with mainstream deep learning methods, PhySR simultaneously achieves recovery of the main structures of extended sources and reconstruction of small-scale details in mixed sources under the strong-degradation scenario, demonstrating better structural fidelity, detail-recovery capability, and reconstruction stability (Fig.~\ref{fig:dl4x}). For the extended-source cases in Figs.~\ref{fig:dl4x}(a) and \ref{fig:dl4x}(b), SR3, DRCT, and HAT exhibit varying degrees of missing and blurred local details. SwinIR primarily recovers the central main region, while the peripheral extended structures are not fully recovered. Although SRFormer recovers the basic morphology of the central source, the local brightness distribution is insufficiently stable and exhibits a certain degree of intensity deviation. By contrast, PhySR more completely recovers the main morphology of the central extended source, preserves continuous and clear edge contours, and retains local details relatively well. For the mixed-source cases in Figs.~\ref{fig:dl4x}(c) and \ref{fig:dl4x}(d), SR3, DRCT, and HAT do not clearly represent the surrounding weak sources and local details, and some small-scale structures are blurred or missing. SwinIR has limited capability for recovering small-scale weak sources, and a certain degree of spurious response remains in the background regions. Although SRFormer recovers some local structures, the edges of the details are overly smooth, and the small-scale structures are insufficiently distinguishable. By contrast, while recovering the central source, PhySR more clearly reconstructs the surrounding weak sources and fine small-scale structures, preserves more complete local details, and effectively suppresses spurious background responses.

\insertwidefigure
{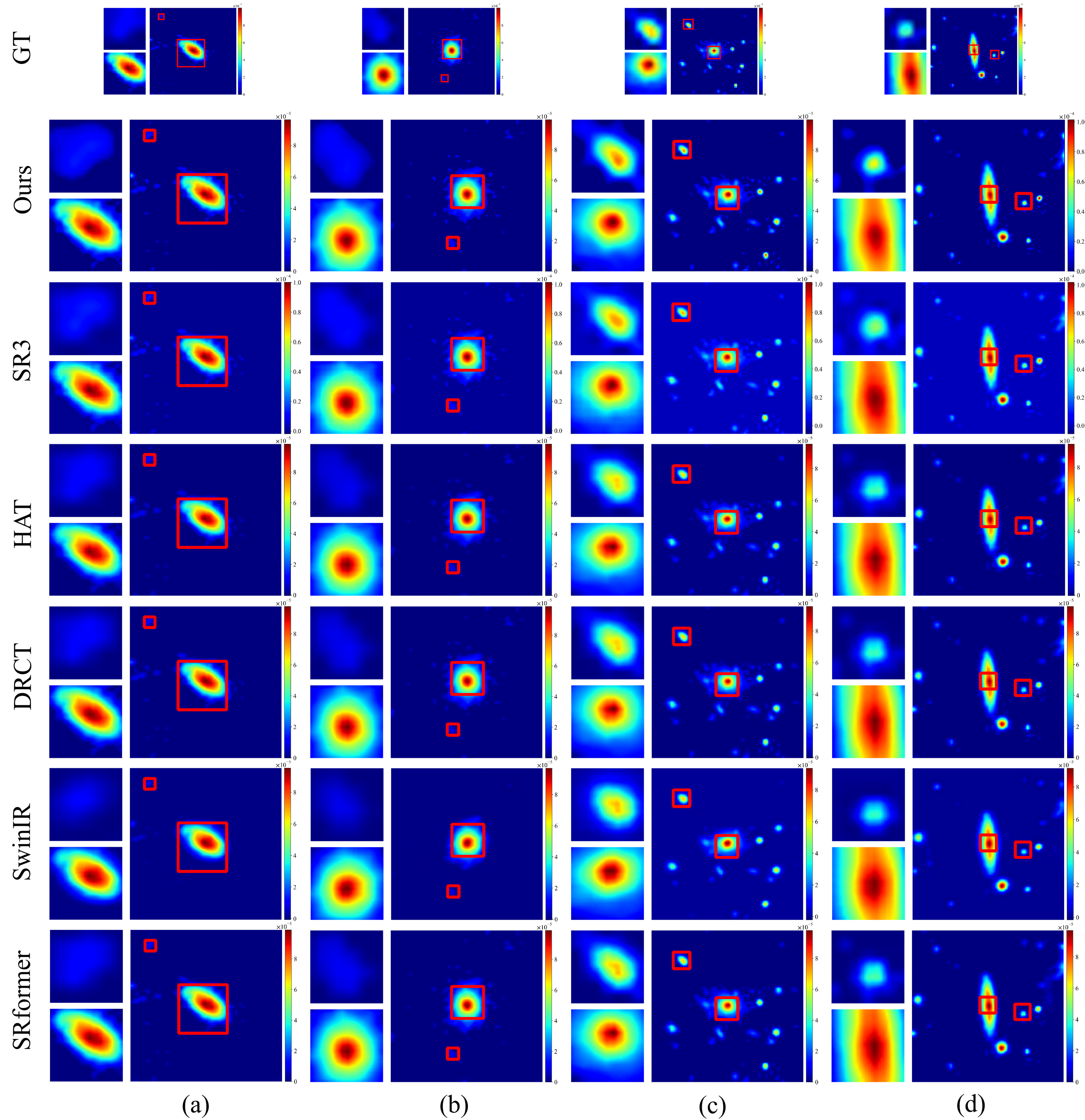}
{0.86\textwidth}
{10.0cm}
{Comparison of 4× super-resolution reconstruction results obtained with PhySR and deep learning methods: panels (a) and (b) show extended sources, whereas panels (c) and (d) show mixed sources; the clean images have a resolution of 64×64 pixels, whereas the reconstructed images produced by the different methods have a resolution of 256×256 pixels; the red boxes indicate the large-scale morphology of the central source (large box) and the fine small-scale details (small box), respectively.}
{fig:dl4x}

Under the extreme-degradation scenario in the 8× super-resolution task, PhySR still maintains relatively stable reconstruction performance and demonstrates stronger capabilities for structural preservation and detail recovery than the other deep learning methods. Figs.~\ref{fig:dl8x}(a) and \ref{fig:dl8x}(b) show extended-source cases, whereas Figs.~\ref{fig:dl8x}(c) and \ref{fig:dl8x}(d) show mixed-source cases. For extended sources, although the different deep learning methods recover the approximate contours of the central source, pixelation remains at the source center, and local details are incompletely recovered. By contrast, PhySR recovers the central extended-source structure relatively completely, with a clear and continuous source center, while also achieving relatively complete recovery of local details. For mixed sources, the other deep learning methods do not sufficiently recover the surrounding weak sources and small-scale structures, and some details are incompletely recovered and exhibit a discretized appearance. While recovering the central source, PhySR reconstructs the small-scale structures and local details more clearly, demonstrating a better capability for detail recovery. Overall, PhySR still maintains relatively stable reconstruction performance under the extreme-degradation scenario in the 8× super-resolution task, demonstrating stronger capabilities for structural preservation and detail recovery.

\insertwidefigure
{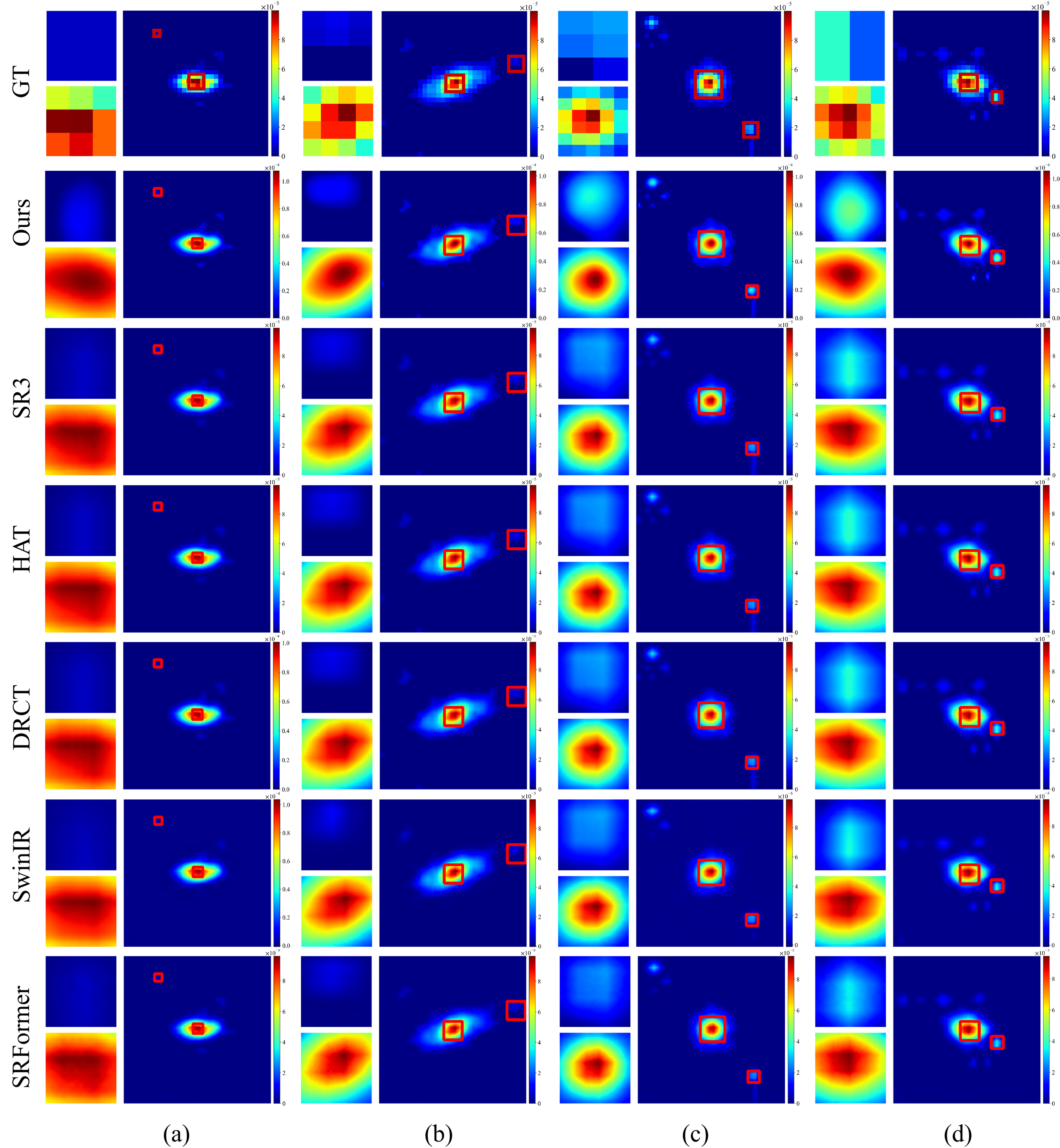}
{0.86\textwidth}
{10.0cm}
{Comparison of 8× super-resolution reconstruction results obtained with PhySR and deep learning methods: panels (a) and (b) show extended sources, whereas panels (c) and (d) show mixed sources; the clean images have a resolution of 32×32 pixels, whereas the reconstructed images produced by the different methods have a resolution of 256×256 pixels; the red boxes indicate the large-scale morphology of the central source (large box) and the fine small-scale details (small box), respectively.}
{fig:dl8x}

Table~\ref{tab:deep_learning} presents a comparison of the mean image-domain metric values of PhySR and the deep learning methods in the 4× and 8× super-resolution tasks. PhySR maintains superior reconstruction performance compared with the other deep learning methods in both the 4× and 8× super-resolution tasks. In the 4× super-resolution task, PhySR achieves PSNR, SSIM, and RMSE values of 44.65 dB, 0.9940, and 0.0065, respectively, attaining the best mean values for all three metrics. In the 8× super-resolution task, PhySR achieves PSNR, SSIM, and RMSE values of 39.19 dB, 0.9700, and 0.0138, respectively, remaining at relatively high levels. By contrast, the metric values of the other deep learning methods decrease markedly, with the best-performing comparison method achieving only 35.55dB, 0.9659, and 0.0178 for the three metrics, respectively, and thus still exhibiting a clear performance gap relative to PhySR. These results demonstrate that PhySR exhibits better robustness, error-control capability, and reconstruction stability under both strong- and extreme-degradation scenarios.

To further evaluate the physical consistency and observational reliability of the PhySR reconstruction results, we present a comparison of observation-domain metrics between the proposed method and the deep learning methods. The results in Table~\ref{tab:deep_observation} show that the values of the observation-domain consistency metrics are relatively close across the deep learning methods and 


\begin{table*}[!t]
\centering

\begin{minipage}[t]{0.455\textwidth}
\vspace{0pt}

\captionof{table}{Comparison of image-domain metrics between PhySR and deep learning methods.}
\label{tab:deep_learning}

\centering

{\fontsize{7.0}{7.8}\selectfont
\setlength{\tabcolsep}{2.6pt}
\renewcommand{\arraystretch}{1.04}

\begin{tabular*}{0.98\linewidth}
{@{\extracolsep{\fill}}llccc@{}}
\hline\hline
Scale & Model & PSNR & SSIM & RMSE \\
\hline

\multirow{6}{*}{$\times4$}
 & SR3      & 38.05 & 0.8549 & 0.0124 \\
 & HAT      & 39.47 & 0.9786 & 0.0116 \\
 & DRCT     & 40.42 & 0.9816 & 0.0103 \\
 & SwinIR   & 35.09 & 0.8356 & 0.0184 \\
 & SRFormer & 37.72 & 0.9781 & 0.0144 \\
 & \textbf{PhySR}
 & \textbf{44.65}
 & \textbf{0.9940}
 & \textbf{0.0065} \\
\hline

\multirow{6}{*}{$\times8$}
 & SR3      & 35.55 & 0.9650 & 0.0178 \\
 & HAT      & 35.55 & 0.9659 & 0.0179 \\
 & DRCT     & 35.18 & 0.9612 & 0.0186 \\
 & SwinIR   & 33.77 & 0.9160 & 0.0224 \\
 & SRFormer & 35.40 & 0.9511 & 0.0191 \\
 & \textbf{PhySR}
 & \textbf{39.19}
 & \textbf{0.9700}
 & \textbf{0.0138} \\
\hline
\end{tabular*}

}
\end{minipage}
\hfill
\begin{minipage}[t]{0.525\textwidth}
\vspace{0pt}

\captionof{table}{Comparison of observation-domain metrics between PhySR and deep learning methods.}
\label{tab:deep_observation}

\centering

{\fontsize{7.0}{7.8}\selectfont
\setlength{\tabcolsep}{1.15pt}
\renewcommand{\arraystretch}{1.04}

\begin{tabular*}{0.98\linewidth}
{@{\extracolsep{\fill}}llcccc@{}}
\hline\hline
Scale & Model & Obs-RMSE & Obs-MAE & RRE & $\sigma_R$ \\
\hline

\multirow{6}{*}{$\times4$}
 & SR3      & 4.52e-06 & 3.17e-06 & 4.34e-01 & 4.10e-06 \\
 & HAT      & 4.52e-06 & 3.15e-06 & 4.17e-01 & 4.13e-06 \\
 & DRCT
 & 4.77e-06
 & 3.16e-06
 & \textbf{4.16e-01}
 & \textbf{4.02e-06} \\
 & SwinIR
 & 4.50e-06
 & \textbf{3.11e-06}
 & 4.17e-01
 & 4.05e-06 \\
 & SRFormer
 & 4.51e-06
 & 3.24e-06
 & 4.18e-01
 & 4.04e-06 \\
 & \textbf{PhySR}
 & \textbf{4.48e-06}
 & 3.15e-06
 & \textbf{4.16e-01}
 & \textbf{4.02e-06} \\
\hline

\multirow{6}{*}{$\times8$}
 & SR3
 & 6.75e-06
 & 4.63e-06
 & 5.97e-01
 & 6.71e-06 \\
 & HAT
 & 6.56e-06
 & 4.53e-06
 & 5.95e-01
 & \textbf{6.40e-06} \\
 & DRCT
 & 6.68e-06
 & \textbf{4.51e-06}
 & 5.82e-01
 & 6.41e-06 \\
 & SwinIR
 & \textbf{6.54e-06}
 & 4.54e-06
 & \textbf{5.81e-01}
 & 6.55e-06 \\
 & SRFormer
 & 6.63e-06
 & 4.72e-06
 & 5.89e-01
 & 6.49e-06 \\
 & \textbf{PhySR}
 & 6.56e-06
 & \textbf{4.51e-06}
 & 5.93e-01
 & 6.52e-06 \\
\hline
\end{tabular*}

}
\end{minipage}

\end{table*}

\FloatBarrier
\noindent
  generally fall within a reasonable error range, indicating that the reconstruction results of the different models retain a certain degree of consistency with the original dirty images after being degraded back to the observation domain. Among them, PhySR maintains stable and relatively favorable observation-domain error values in both the 4× and 8× tasks, indicating good physical consistency and observational reliability. It should be noted that observation-domain metrics primarily measure the residual level at the observation scale and cannot fully reflect the structural integrity and detail quality of high-resolution reconstructed images. Therefore, although some deep learning methods achieve better results for individual observation-domain metrics, their local reconstruction results still exhibit problems such as blurred edges, insufficient recovery of weak sources, and missing small-scale details, whereas PhySR demonstrates more balanced performance in both observation-domain consistency and image-structure recovery.

\begin{figure*}[!tb]
\centering
\includegraphics[
    width=0.74\textwidth,
    height=0.76\textheight,
    keepaspectratio
]{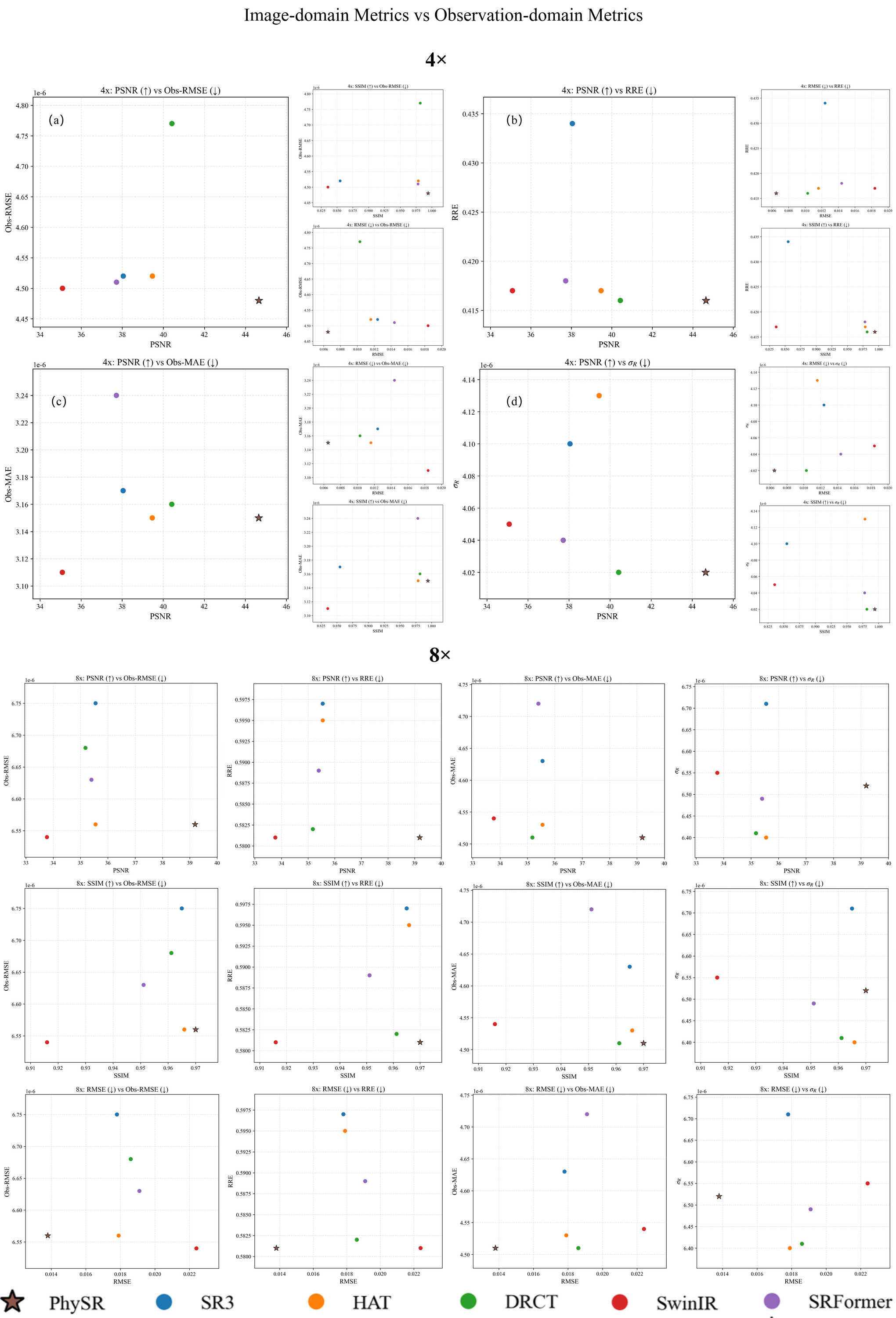}
\caption{Scatterplot matrix of the image-domain metrics and observation-domain consistency metrics. The upper and lower halves show the results for the 4× and 8× super-resolution tasks, respectively. The horizontal axes represent the image-domain metrics, including PSNR, SSIM, and RMSE, whereas the vertical axes represent the observation-domain consistency metrics, including Obs-RMSE, Obs-MAE, RRE, and \(\sigma_{\text{R}}\). For PSNR and SSIM, higher values indicate better image reconstruction quality, and an ideal model should be located toward the far right of the horizontal axis; for RMSE and all observation-domain consistency metrics, lower values indicate smaller errors, and an ideal model should be located toward the bottom of the vertical axis. Therefore, an ideal model should exhibit low observation-domain errors while maintaining high PSNR and SSIM values and a low RMSE value.}
\label{fig:scatter}
\end{figure*}

To analyze the model's ability to balance image-domain reconstruction quality and observation-domain consistency, this work presents scatterplot matrices of the image-domain metrics and observation-domain consistency metrics (Fig.~\ref{fig:scatter}). The upper and lower parts of the figure correspond to the 4× and 8× super-resolution tasks, respectively; different colors denote different models, and star-shaped markers denote PhySR. The horizontal axes represent the image-domain metrics PSNR, SSIM, and RMSE, whereas the vertical axes represent the observation-domain consistency metrics Obs-RMSE, Obs-MAE, RRE, and \(\sigma_{\text{R}}\). An ideal model should exhibit favorable performance in both the image domain and the observation domain. Taking the 4× super-resolution task as an example, when the horizontal-axis metric is PSNR or SSIM, an ideal model should be located in the lower-right region of the figure, indicating both high image reconstruction quality and low observation-domain residuals. When the horizontal-axis metric is RMSE, an ideal model should be located in the lower-left region, indicating low error levels in both the image domain and the observation domain. As shown in Figs.~\ref{fig:scatter}(a), \ref{fig:scatter}(b), \ref{fig:scatter}(c), and \ref{fig:scatter}(d), PhySR is closer to the ideal region in all the corresponding scatterplots, indicating that it achieves favorable image-domain metric values while maintaining observation-domain errors within a relatively low range. The scatter distributions for the remaining observation-domain metrics and the 8× super-resolution task exhibit similar patterns, further demonstrating that PhySR can balance image reconstruction quality and observation-domain physical consistency under strong- and extreme-degradation scenarios, thereby achieving better overall reconstruction performance and stability.

\section{Conclusion}\label{sec:conclusion}

This work proposes PhySR, an end-to-end super-resolution reconstruction model based on a physics-informed neural network (PINN). Using U-Net as its backbone, the proposed method explicitly embeds the physical priors of the primary beam response and PSF convolution in radio synthesis imaging into the network training process and combines a dynamic cascaded upsampling module with a multiscale feature residual module, thereby enabling unified modeling of the elimination of the coupling effects between the primary and synthesized beams and image super-resolution reconstruction. Compared with existing general-purpose methods that rely on stepwise processing and data-driven deep learning methods for super-resolution, PhySR can directly recover high-fidelity images from low-resolution dirty images within an end-to-end framework, improving image reconstruction quality while ensuring physical consistency.

The experimental results show that PhySR achieves stable and superior reconstruction performance across different degradation levels. The ablation experiments validate the effectiveness of the physical consistency constraint, the dynamic cascaded upsampling module, and the multiscale feature residual module. The joint use of these three components effectively reduces reconstruction errors, improves source-structure recovery, and enhances the physical consistency of the reconstructed results after they are degraded back to the observation domain. The PhySR model achieves good reconstruction results for both source types; compared with existing general-purpose methods, it attains PSNR, SSIM, and RMSE values of 44.65 dB, 0.9940, and 0.0065, respectively, in the 4× super-resolution task, and 39.19 dB, 0.9700, and 0.0138, respectively, in the 8× task, clearly outperforming the comparison methods, including Hg-CLEAN+PBC+BIU, MS-CLEAN+PBC+BIU, and ASP-CLEAN+PBC+BIU, thereby demonstrating the significant advantages of the proposed method in eliminating beam effects, preserving structural fidelity, and controlling reconstruction errors.

Compared with mainstream deep learning models for super-resolution, PhySR achieves PSNR, SSIM, and RMSE values of 44.65 dB, 0.9940, and 0.0065, respectively, in the 4× task, outperforming SR3, HAT, DRCT, SwinIR, and SRFormer across all three metrics. Among the comparison methods, DRCT achieves relatively high PSNR and SSIM values of 40.42 dB and 0.9816, respectively, both of which remain lower than those achieved by PhySR. In the 8× task, PhySR still achieves PSNR, SSIM, and RMSE values of 39.19 dB, 0.9700, and 0.0138, respectively, outperforming the corresponding values of 35.55 dB, 0.9659, and 0.0179 achieved by HAT, the best-performing comparison method. These results demonstrate that the proposed method exhibits better image fidelity, structural recovery capability, and error-control capability under both strong- and extreme-degradation scenarios.

Moreover, this work introduces observation-domain consistency metrics to evaluate the reliability of the reconstruction results from the perspective of physical observation. The experimental results show that the deep learning methods exhibit generally similar observation-domain metric values, all of which fall within a reasonable error range. In the 4× task, PhySR achieves Obs-RMSE, Obs-MAE, RRE, and \(\sigma_{\text{R}}\) values of 4.48e-06, 3.15e-06, 4.16e-01, and 4.02e-06, respectively; in the 8× task, the corresponding values are 6.56e-06, 4.51e-06,5.93e-01, and 6.52e-06, respectively. Although some deep learning methods achieve near-optimal results for individual observation-domain metrics, the image-domain metrics, the local reconstruction results for extended and mixed sources, and the scatterplot matrix analysis collectively show that PhySR achieves higher-quality image reconstruction results while maintaining good observation-domain consistency, thereby demonstrating better image fidelity, structural recovery capability, and physical reliability.

This work primarily uses simulated SKA-Mid observational data to validate the feasibility and effectiveness of PhySR in eliminating the coupling effects between the primary and synthesized beams and performing super-resolution reconstruction. Future research can be pursued in the following directions: (1) validation and transfer studies using real radio observational data, such as data from MeerKAT and LOFAR, to further examine the model's generalization capability and scope of applicability; and (2) extension of the proposed method from the current super-resolution reconstruction task to more complex applications in radio synthesis imaging, such as wide-field imaging, fine-structure recovery of weak sources, and joint deconvolution and super-resolution reconstruction, thereby expanding its application potential in data processing for next-generation radio telescopes.

\FloatBarrier


\begin{acknowledgements}
The work was partially supported by the Guizhou Provincial Leading Talent Development Program (No.KJLYRC-[2026]007), the Guizhou Provincial Basic Research Program (Natural Science) (No. QNA[2026]005), the National Natural Science Foundation of China (No.12273007), the Guizhou Provincial Program on Commercialization of Scientific and Technological Achievements (No. KJFZ[2025]014), the Guizhou Provincial Excellent Young Science and Technology Talent Program (No. YQK[2023]006), the National Key R\&D Program of China (No.2022YFE0133700), the National Natural Science Foundation of China (No.11963003), the National SKA Program of China (No.2020SKA0110300), the Guizhou Provincial Basic Research Program (Natural Science) (No.ZK[2022]143).
\end{acknowledgements}

\bibliographystyle{bibtex/aa}
\bibliography{references}

\end{document}